\documentclass[]{aastex62}

\usepackage{graphicx}
\usepackage{float}
\usepackage[caption=false]{subfig}
\usepackage{enumitem}
\usepackage{amsmath}
\usepackage{url}
\newcommand{\degree}{^\circ}

\newcommand{\combinedNoise}{9} 
\defcitealias{schlawin2020jwstNoiseFloorI}{Paper I}
\newcommand{\paperI}{\citetalias{schlawin2020jwstNoiseFloorI}}

\submitjournal{AJ}
\received{October 7, 2020}
\revised{December 8, 2020}
\accepted{January 4, 2021}

\shorttitle{NIRCam Lab Stability Studies}
\shortauthors{Schlawin et al.}

\begin{document}

\title{JWST Noise Floor II: Systematic Error Sources in JWST NIRCam Time Series}

\correspondingauthor{Everett Schlawin}
\email{eas342 AT EMAIL Dot Arizona .edu}

\author[0000-0001-8291-6490]{Everett Schlawin}
\affiliation{Steward Observatory \\
933 North Cherry Avenue \\
Tucson, AZ 85721, USA}

\author{Jarron Leisenring}
\affiliation{Steward Observatory \\
933 North Cherry Avenue \\
Tucson, AZ 85721, USA}

\author{Michael W. McElwain}
\affiliation{NASA Goddard Space Flight Center \\
Exoplanets and Stellar Astrophysics Laboratory \\
Greenbelt, MD 20771, USA}

\author{Karl Misselt}
\affiliation{Steward Observatory \\
933 North Cherry Avenue \\
Tucson, AZ 85721, USA}

\author{Kenneth Don}
\affiliation{Steward Observatory \\
933 North Cherry Avenue \\
Tucson, AZ 85721, USA}

\author{Thomas P. Greene}
\affiliation{NASA Ames Research Center \\
Space Science and Astrobiology Division \\
Moffett Field, CA 94035, USA}

\author{Thomas Beatty}
\affiliation{Steward Observatory \\
933 North Cherry Avenue \\
Tucson, AZ 85721, USA}

\author{Nikolay Nikolov}
\affiliation{Space Telescope Science Institute\\
3700 San Martin Dr \\
Baltimore, MD 21218, USA}

\author{Douglas Kelly}
\affiliation{Steward Observatory \\
933 North Cherry Avenue \\
Tucson, AZ 85721, USA}

\author{Marcia Rieke}
\affiliation{Steward Observatory \\
933 North Cherry Avenue \\
Tucson, AZ 85721, USA}



\begin{abstract}

JWST holds great promise in characterizing atmospheres of transiting exoplanets, potentially providing insights into Earth-sized planets within the habitable zones of M dwarf host stars if  photon-limited performance can be achieved.
Here, we discuss the systematic error sources that are expected to be present in grism time series observations with the NIRCam instrument.
We find that pointing jitter and high gain antenna moves on top of the detectors' subpixel crosshatch patterns will produce relatively small variations (less than 6 parts per million, ppm).
The time-dependent aperture losses due to thermal instabilities in the optics can also be kept to below 2 ppm.
To achieve these low noise \replaced{sources}{values}, it is important to employ a sufficiently large (more than 1.1 arcseconds) extraction aperture.
Persistence due to charge trapping will have a minor (less than 3 ppm) effect on time series 20 minutes into an exposure and is expected to play a much smaller role than it does for the HST WFC3 detectors.
We expect \added{detector} temperature fluctuations to \replaced{be}{contribute} less than 3 ppm.
In total, our estimated noise floor from known systematic error sources is only \combinedNoise\ ppm per visit.
We do however urge caution as unknown systematic error sources could be present in flight and will only be measurable on astrophysical sources like quiescent stars.
We find that reciprocity failure may introduce a perennial instrument offset at the 40 ppm level, so corrections may be needed when stitching together a multi-instrument multi-observatory spectrum over wide wavelength ranges.
\end{abstract}




\section{Introduction} \label{sec:intro}

The James Webb Space Telescope (JWST) will be a powerful new tool to characterize exoplanets, especially on transiting systems.
The transmission and emission spectra of planets will permit new avenues to measure the composition, temperature profiles and maps in a variety of planets \citep[e.g.][]{beichman2014pasp,greene2016jwst_trans,howe2017informationJWST,barstow2015jwstSystematics,schlawin2018JWSTforecasts}.

It is critical, however, that JWST perform close to its photon noise limit in order to fully characterize \deleted{the} exoplanet atmospheres with small scale heights relative their host stars' radii.
This is especially true if JWST is to be used to search for biosignatures in habitable-zone terrestrial planets, such as the TRAPPIST-1 system \citep[e.g.][]{barstow2016trappist1habitable,krissansen-totton2018trappist1eJWST,pidhorodetska2020waterTRAPPIST-1e}.
While there would be enough photons to detect an ozone feature in 30 transits of TRAPPIST-1 d \citep{barstow2016trappist1habitable}, there could be systematic errors that complicate the measurement.
Systematic errors can come from astrophysical sources such as un-occulted starspots \citep[e.g.][]{rackham2018transitSourceEffect}.
They can also occur due to observatory and instrument artifacts from JWST itself.

\added{We focus here on time series applications of JWST NIRCam, which also have applicable lessons for other near-infrared instruments that use HgCdTe detectors on JWST or ground-based observatories.
NIRCam is the Near-Infrared Camera on JWST, capable of imaging two 2\arcmin\ $\times$ 2\arcmin\ fields of view simultaneously with a short wavelength filter (between 0.6~\micron\ and 2.3~\micron) and a long wavelength filter (between 2.4~\micron\ and 5.0~\micron) \citep{rieke2005nircamSPIE}.
NIRCam has a grism pupil element that can be used on the long wavelength channel for spectroscopy \citep{greene2016wfss,greene2017jatisNIRCam}.
In the grism time series mode, the short wavelength channel can be used with weak lens imaging or with a dispersed Hartmann sensor being discussed as a science enhancement \citep{schlawin2017dhs}.

The time series observation modes of NIRCam can capture changes in the flux of science target(s) to capture astrophysical variable phenomena, such as planets eclipsing/transiting their host stars, rotational modulations or pulsation modes in stars.
The time series modes point the telescope as stably as possible at the target(s) without dithering, re-pointing or moving mechanisms in order to measure the flux as precisely as possible.
This is in contrast to other standard JWST modes that dither (move) the telescope to mask bad pixels, reduce pixel sampling errors and reduce flat field errors.
With time series modes, the goal is not absolute flux accuracy or un-pixelated images, but rather precision measurements of flux changes.
The resulting data from time series modes consists of a series of integrations of the same field where it is possible to study time-variations in the pixels.

NIRCam has 4 short wavelength detectors and 1 one long wavelength detector over each of its 2\arcmin\ $\times$ 2\arcmin\ fields of view, totaling 10 detectors in all.
The field of views are split into the ``A'' side and ``B'' side and numbered A1 through A4 and B1 through B4 for the short wavelength channels and A5 and B5 for the long wavelength channels.
Time series imaging uses 5 of the detectors on the ``B'' side for high precision photometry.
The grism time series uses the grism pupil element to disperse the spectrum at a resolution of $\sim1100$ to $1700$ \citep{greene2017jatisNIRCam} on the A5 detector while imaging simultaneously with the A1 through A4 detectors.
We expect the grism time series to be the most frequently employed time series mode on NIRCam and the science to come from the spectrum, so we default most calculations in this paper to the A5 detector.
}

\citet{suissa2020oceanEarthsMStars} find that JWST's noise floor is a critical factor in detecting H$_2$O features of Earth-sized ocean-bearing worlds around M Dwarfs.
For the Transiting Exoplanet Survey Satellite (TESS) input catalog, it is not possible to detect water vapor in water-bearing worlds if the absolute noise floor exceeds 5 ppm.
However, if the noise floor is 3 ppm, there are about 8 exoplanet systems for which JWST could detect 2.7\micron\ water vapor features in less than 100 hours of accumulated exposure time.
Thus, the noise floor is a critical parameter in future searches for habitable conditions and the presence of atmospheres on Earth-sized worlds.

In \citet{schlawin2020jwstNoiseFloorI}, hereafter \paperI, we discussed the random errors that increase the noise of JWST NIRCam time series observations above the photon noise.
\deleted{We showed that 1/f noise will be the dominant random error that adds to lightcurves and can be larger than the photon noise when uncorrected for some aperture sizes.
Fortunately, 1/f noise can be mitigated and reduced to below photon errors with background subtraction and a covariance-weighting spectral extraction scheme.
Alternatively, the GRISMC element, if implemented, can be used to reduce the 1/f noise contribution because it disperses spectra perpendicular to the noise correlations.}
Here, we discuss the systematic errors that can increase the noise in JWST time series, again with a focus on the NIRCam instrument.
\added{We summarize important results from \paperI\ in the end of Section \ref{sec:knownEffects}}
The NIRSpec and NIRISS time series modes may have similar systematic errors because they have the same kinds of detectors and similar telescope pointing jitter.
However, there will be some differences because NIRSpec bright object time series uses a slit \citep{ferruit2014transitingPNIRSpec} and NIRISS single object slitless spectroscopy has a unique cylindrical lens to spread the light in the cross-dispersion direction \citep{doyon2012NIRISSFGS,albert2014nirissGratingPrism}.

While photon noise and the random errors discussed in \paperI\ will fall as $1/\sqrt{N}$ for $N$ integrations, systematic errors may fall more slowly with $N$.
If the systematic errors cause correlations between integrations, they will fall less rapidly than $1/\sqrt{N}$ or may not decrease at all with $N$ if integrations are perfectly correlated.
Systematic errors can introduce non-astrophysical time-dependent behavior to JWST lightcurves or produce a deterministic offset between the true signal and the measured one.
Often, systematics can be functions of other variables such as the telescope pointing, wavefront errors in the JWST optics, a detector's temperature or the charge trap state in the detector.

Section \ref{sec:knownEffects} lists the known systematic effects that can degrade the precision of time series measurements compared to the photon noise limit.
Sections \ref{sec:subpxCrosshatch} through \ref{sec:recipFailure} provide simulations and analysis of these effects to provide some quantitative estimates on how much they could potentially contribute to JWST NIRCam time series.
We conclude in Section \ref{sec:Conclusion} that the known systematic errors we study contribute $\sim$\combinedNoise\ ppm or smaller effects, but more tests in flight will be necessary to verify performance and develop mitigation strategies.

\section{List of Known Systematic Effects}\label{sec:knownEffects}

In \paperI, we described the random effects in NIRCam time series that increase the scatter of lightcurves.
In addition to these random effects, there are also systematic effects that can produce correlated noise between integrations due to other variables such as temperature and telescope pointing.
Here, we consider systematic effects that can introduce longer timescale variations and impact the transit or eclipse depths of planets.

\begin{enumerate}[noitemsep]
	\item \textbf{Intrapixel sensitivity:} Telescope pointing drifts and jitter can impact the amount of flux measured if the response within pixels is not perfectly uniform.
	This has a particularly strong impact on the time series from the Spitzer IRAC instrument \citep{ingalls2016spitzerRepeatability}.\label{it:intrapixelSens}
	It can also appear on JWST detectors because the sub-pixel flat field is not uniform.
	Crosshatching patterns on HgCdTe detectors, which all JWST near-infrared instruments use, exist at the sub-pixel level  \citep{shapiro2018crosshatch,ninan2019crosshatchHPF}.
	The intrapixel sensitivity is more pronounced when the point spread function (PSF) is under-sampled and there are pointing variations, such as with the Spitzer IRAC imager.
	\item \textbf{Variable Aperture Losses:} A photometric or spectroscopic lightcurve will vary in time if the amount of flux from a star within its assigned extraction aperture varies.
	JWST is expected to have wavefront error variations due to slight motions of the mirror surfaces with thermal expansion and contraction that result in subtle changes in JWST's PSF.
	If the amount of flux contained within an extraction aperture (sometimes referred to as encircled energy) varies on hour-long timescales across a transit or eclipse, this can cause spurious variations in planet transit or eclipse depth measurements.\label{it:apLoss}
	\item \textbf{Detector Temperature Fluctuations:} The JWST detectors are sensitive to temperature changes on the focal plane arrays. For example, laboratory tests show that 100 mK temperature fluctuations can result in $\sim$80$~e^-$ changes in the detector bias offsets that are not corrected by reference pixels \citep{hall2005jwstArrays}. NIRCam detectors are actively thermally controlled to keep temperature fluctuations to $\lesssim$ 1 mK.
	However, during cryovac ground testing, the Long Wavelength (LW) A detector (A5) exhibited $\sim$20 min oscillations with an amplitude of $\sim$15-20 mK when using the primary Temperature Monitor Control (TMC).
	Switching to the redundant TMC, an independent heater and Cernox temperature sensor for A5, showed no observable oscillations; this configuration will be employed in orbit.
	If necessary, the effect of temperature fluctuations can be calibrated to $\pm 1 e^-$ for excursions less than $\pm$ 50 mK \citep{hall2005jwstArrays}.
	\item \textbf{Charge Trapping} A major correlated error source on the Hubble Space Telescope's Wide Field Camera 3 (HST's WFC3) is the ramp effect \citep{berta2012flat_gj1214}.
	The ramp effect is due to charge trapped within the detector's depletion region, initially lowering the measured flux, and then released after detector reset, thereby increasing the measured flux most prominently at the beginning of an integration \citep{zhou2017chargeTrap}.
	JWST's NIRCam instrument has very low\deleted{er} rates of persistence ($<$1 DN/sec) at 100 seconds after saturation on all detectors \citep{leisenring2016persistence} and JWST will observe continuously without Earth occultations experienced by HST, so it is expected that charge trapping will be a minor concern for JWST's HgCdTe detectors.
	This allows pixels to reach a steady state, assuming that the pointing jitter is within a pixel, as expected for JWST.
	\item \textbf{Reciprocity Failure:} The response of a HgCdTe detector can change with the brightness of a source, which is called reciprocity failure \citep{biesiadzinski2011reciprocityFailure}.
	The brightness of a star and planet system does not change significantly with time, so it will not significantly affect the {\it precision} of a lightcurve, but could affect {\it accuracy} because the filter bandpass and stellar spectral energy distribution change the count rates with wavelength.
	It is possible that reciprocity failure could introduce a systematic offset, slope or curvature that depends on the count rate of the detector.
	\item \textbf{Ghost Images and Contaminating Light:} There are 17 optical surfaces in the NIRCam long wavelength channel that can create spurious reflections or ghost images.
	Ground-based testing revealed a ring-like ghost on grism images with up to 1.5\% the value as the average flux in a grism aperture.
	This is a second order effect because it contaminates the measured flux by 1.5\% the difference in transit depth.
	So, for a 100 ppm variation in transit depth with wavelength, we expect about 1.5 ppm differences from the peak part of ghost images.
	This would be present in all images, so it could produce a systematic offset but not a time-variable signature.
\end{enumerate}

\added{
The systematic errors discussed here should be compared with the random error sources discussed in \paperI.
We found in \paperI\ that detector noise sources can exceed the photon noise for short integrations and thus can affect science on bright targets.
The main challenge to reducing the detector noise is that it highly correlated along pixels in the fast-read direction parallel to the spectrum.
When averaging together the flux from many pixels in wavelength, the read noise stays constant, rather than falling as the $\sqrt{N_{px}}$ where $N_{px}$ is the number of pixels, as would occur if each pixel was independent.
For the shortest integrations where there are only two read samples, the 1/f detector noise can exceed 830 ppm in bins that are 0.17~\micron-wide and 14 spatial-pixels tall, compared to 390 ppm for the photon noise.

Fortunately, the detector read noise can be reduced by adjusting the observing and extraction strategy.
Operationally, a GRISMC mode is being discussed that disperses the spectrum perpendicular to the 1/f noise correlations, permitting them to be dramatically reduced with background subtraction.
The read noise can be mitigated by using a very narrow (3 pixel or 0.19\arcsec) aperture or a co-variance weighting scheme that takes into account pixel noise correlations.
The co-variance weighting scheme can reduce the contribution due to read noise to 230 ppm per read pair sample.
This very narrow aperture, however, could be more susceptible to Systematic Effect \ref{it:intrapixelSens} (intra-pixel sensitivity) and Systematic Effect \ref{it:apLoss} (variable aperture losses) discussed above.
If many groups are averaged together, this will drop the read noise by a factor of $\sqrt{N_{read}}$, where $N_{read}$ is the number of non-destructive reads.
For a typical $\sim$1 hour transit duration a subarray size of 2048$\times$256 and 5 groups, there are a total of 2290 read samples per pixel.
This will reduce the 230 ppm random noise for 2 read samples by $\sqrt{2290/2}$ or 7 ppm in the transit depth.
We therefore are interested in effects that are of order 7~ppm.
For bright targets that use even smaller subarrays such as the 2048$\times$64 subarray, we are interested in systematic errors that are a factor of about $\sqrt{4}$ smaller or 3.5~ppm.
}

\section{Subpixel Crosshatching}\label{sec:subpxCrosshatch}
\begin{figure}[!hbtp]
\centering
\includegraphics[width=.49\columnwidth]{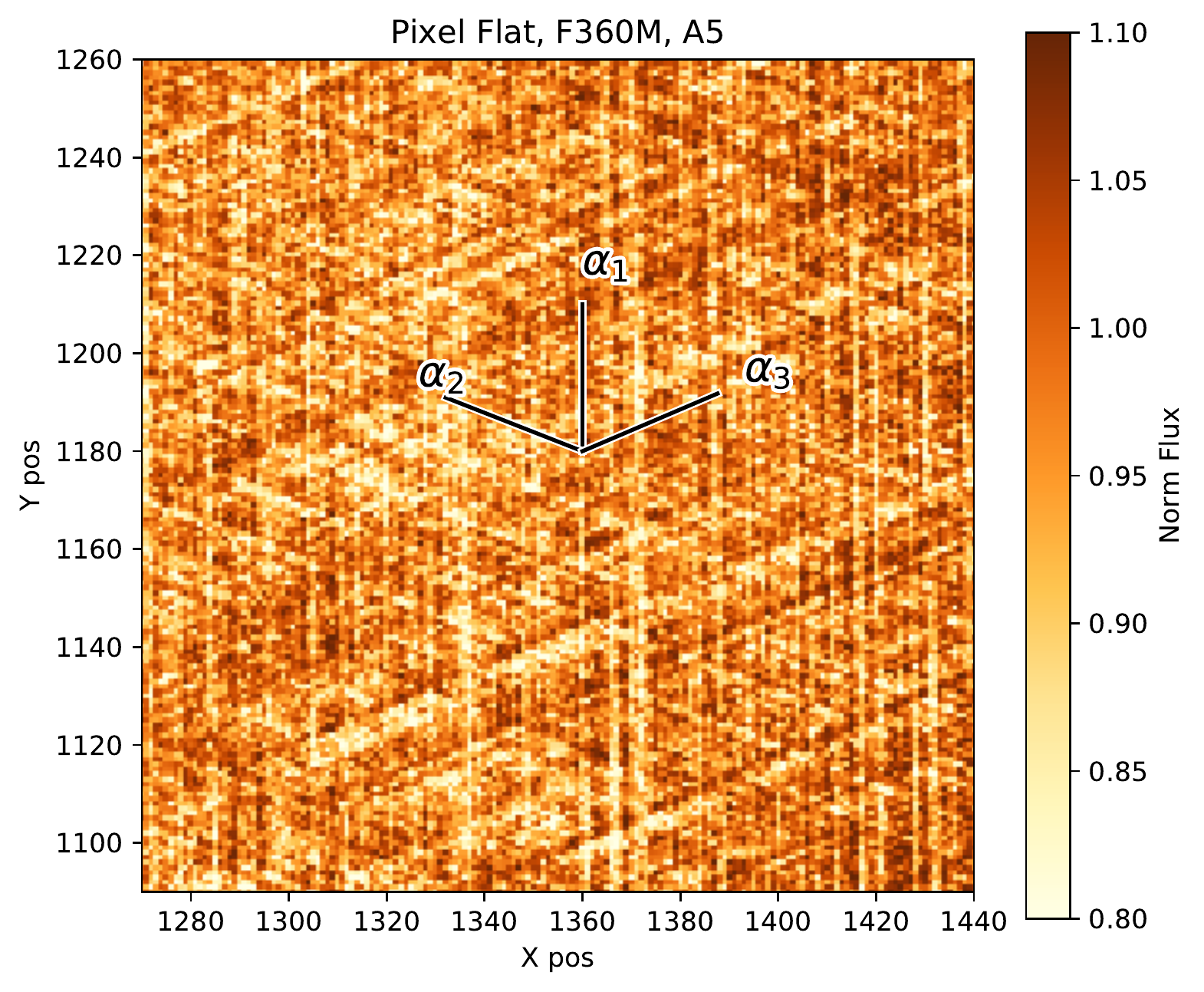}
\caption{
The A5 detector's flat field has a pronounced crosshatch pattern at $\alpha_1 = 90.9 \degree$, $\alpha_2 = 158.6\degree$ and $\alpha_3 =23.1\degree$ CCW from the positive X direction.
The crosshatch pattern is visible in a zoom-in of the pixel flat field and it exists at the sub-pixel level as well.
When the image of a star moves on top of this crosshatch pattern, it will create flux variations.
\deleted{ (top left).
The 2D power spectrum of the flat field (top right) shows a continuum of frequencies aligned with the three crosshatch angles.
Note that the lines in the frequency plane ($\theta_i$) are perpendicular to the crosshatch directions in the image plane ($\alpha_i$).
These three relative crosshatch angles (67.8$\degree$, 67.7$\degree$ and 44.5$\degree$) are similar to a projection of a tetrahedral (zincblende) lattice structure of HgCdTe (bottom, 67.8$\degree$, 67.8$\degree$, 44.4$\degree$).
The green circles represent either Cd or Hg while the red represents Te.}
}\label{fig:crossHatchA5Zoom}
\end{figure}

The NIRCam detectors, most notably the A5 detector used in grism time series, have a crosshatching pattern in their flat field profile.
\added{Figure \ref{fig:crossHatchA5Zoom} shows a zoom-in of the A5 detector's flat field calibration image from Cryogenic Vacuum Testing 3 (CV3) at NASA Goddard.
The crosshatch pattern extends to the subpixel level and thus can cause flux variations when the centroid of an image or spectrum moves, even on a subpixel level.
In appendix \ref{sec:detailedSupxCrosshatchingCharacterization}, we describe in more detail the characterization of the crosshatch pattern and its connection to the crystallographic structure of the HgCdTe detector material.
Observers interested in the direct consequences of subpixel effects, rather than the details of the calculation, can skip to Section \ref{sec:subpxCrossHatchScanResults}.}

We also briefly compare the amplitude of the crosshatch signal in our A5 detector with published measurements of \added{the} crosshatch pattern and intra-pixel sensitivity for other H2RG detectors.
\citet{shapiro2018crosshatch} examine the amplitude of the crosshatch pattern in a candidate detector for the Euclid mission that has particularly strong crosshatch features.
The amplitude of variation in the flat field signal for the candidate Euclid detector is about 1.5 to 2 times the level of the NIRCam A5 detector.
We also note that \citet{hardy2008intrapixelReponseJWST,hardy2014intrapixelResponseJWST} measured the intrapixel response of a detector produced for the Fine Guidance Sensor of JWST.
\citet{hardy2014intrapixelResponseJWST} find that on average, there is no sensitivity across a pixel because the flux is distributed to neighboring pixels.
So long as one uses a large extraction aperture to include neighboring pixels, there is no average intrapixel response function affecting all pixels.
However, there are defects or large drops in the intrapixel map with amplitudes of $\sim$10\% that could result in photometric variations.
The map in \citet{hardy2008intrapixelReponseJWST} of earlier-generation JWST detectors shows long linear features that may be related to crosshatch patterns.
The map in \citet{hardy2014intrapixelResponseJWST} does not show a strong crosshatch pattern but the amplitude of the defects are similar to the crosshatch pattern dips we measure from the NIRCam A5 flat field.
We therefore expect that our analysis of the crosshatch pattern will give similar amplitude effects as the \citet{hardy2014intrapixelResponseJWST} measurements.
On-flight measurements of the subpixel sensitivity of NIRCam will reveal if there similar defect structures at the location of the grism time series field point.

\subsection{Crosshatch Pattern Modeling}\label{sec:crosshatchModeling}
We model the crosshatch pattern in frequency space because it has a smoother pattern as a function of spatial frequency than in spatial coordinates as seen in Figure~\ref{fig:crossHatchA5}.
We divide the 2D Fourier power spectral density by the number of pixels in an input image.
This was experimentally determined to give a power spectral density that does not change with the dimensions of the input image.

We model the two dimensional Fourier power spectral density $f(k_{x,i}',k_{y,i}')$ for one angle $\theta_i$ as
\begin{equation}\label{eq:analyticPSDradial}
f(k_{x,i}',k_{y,i}', a_i, b,c) = a_i \exp{\left(- |k_{x,i}'| / b \right)} \left( \frac{1}{(0.5 c)^2 + k_y'^2} \right),
\end{equation}
where $k_x'$ and $k_y'$ are rotated frequency coordinates in the parallel and perpendicular directions, $a$ is the amplitude, $b$ is the parallel exponential constant and $c$ is the Lorentzian full width at half maximum of the perpendicular dependence.
For each of the three angles, there is a set of rotated coordinates centered at $(k_{x,0},k_{y,0})$= (0,0) following:
\begin{equation}
k_x'(\theta_i) = k_x \cos{\theta_i} + k_y \sin{\theta_i}
\end{equation}
and
\begin{equation}
k_y'(\theta_i) = -k_x  \sin{\theta_i} + k_y \cos{\theta_i}.
\end{equation}
We include three angles so that the total power spectral density is
\begin{equation}\label{eq:fullPSDmodel}
f(k_x,k_y) = \sum_{i=1}^{i=3} f_i(k_{x,i}'(\theta_i),k_{y,i}'(\theta_i),a_i,b,c) + d \exp{\left(-k_r/e_b\right)},
\end{equation}

This model has 8 free parameters: three angles ($\theta_i$), three amplitudes ($a_i$), a joint parallel exponential constant ($b$) and joint Lorentzian width ($c$).
Finally, we include an exponential term that fits the broad azimuthally symmetric background to all the power spectra,
where $d$ is the amplitude of the radial ``background'', $k_r= \sqrt{k_x^2+k_y^2}$ is the radial frequency and $e_b$ is the radial background exponential constant.
We only fit the region of the power spectral density above frequencies of 0.05 px$^{-1}$ to focus on the high frequency component of the flat field where the crosshatch is most prevalent.
The lower spatial frequencies describe broader structures like epoxy voids or illumination gradients that are less relevant for time series observations.

For the F300M filter and A5 detector, we find $\theta_1 = 0.906 \pm 0.001 \degree, \theta_2 = 68.605 \pm 0.001 \degree, \theta_3 = 113.075 \pm 0.001, a_1 = 2.774 \pm 0.002 \times 10^{-7}, a_2 = 1.569 \pm 0.002 \times 10^{-7}, a_3 = 1.744 \pm 0.002 \times 10^{-7}, b=0.345 \pm 0.0005 $ px$^{-1}$, $c=2.05 \pm 0.07 \times 10^{-2}$ px$^{-1}$, $d=6.78 \pm 0.01 10^{-3}$ and $e_b=0.26 \pm 0.01$ px$^{-1}$.
In other words, the separations between the vectors are $67.699 \degree, 67.831 \degree$, and $44.470 \degree$.
The uncertainties were simply derived from the diagonals of the covariance matrix and may be underestimated.

\begin{figure}[!hbtp]
\centering
\includegraphics[width=.99\columnwidth]{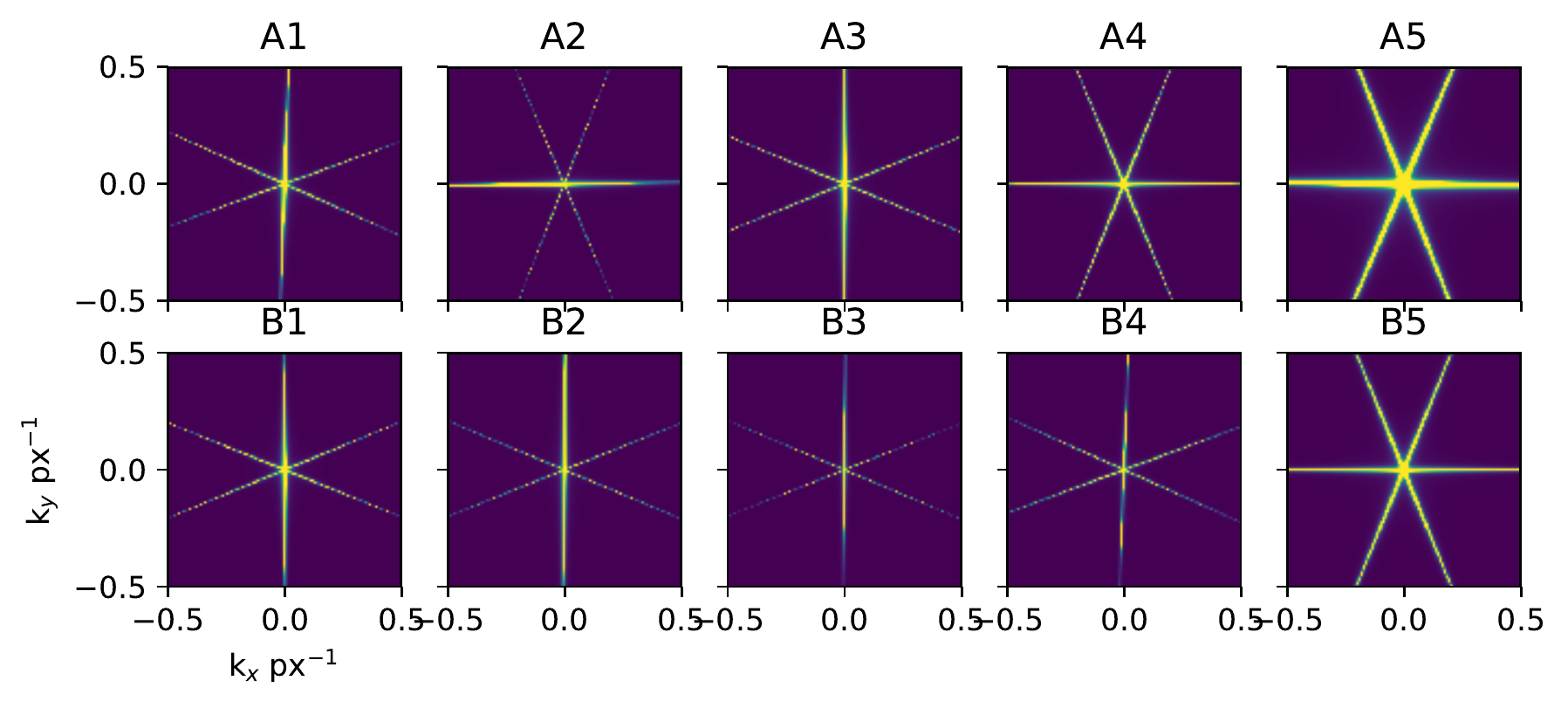}
\caption{
Best-fit power spectra of all the NIRCam detectors show that the long wavelength detectors A5 and B5 show the most pronounced crosshatch patterns.
The primary crosshatch tends to be aligned closely, but not exactly, with either the X or Y axis.
The images have all been normalized to the same scale to show the relative crosshatch amplitudes of different detectors.
}\label{fig:crosshatchModelGallery}
\end{figure}

NIRCam's 10 sensor chip assemblies (SCAs) have very similar crosshatch angles with average separations of 44.5$\degree$ and 67.7 to 67.8$\degree$ as seen in Figure~\ref{fig:crosshatchModelGallery}.
While the absolute orientations relative to pixel axes change, the relative angles of the three crosshatch axes are within 0.15$\degree$ from the average.
They all have one primary crosshatch axis oriented either nearly along the X pixels or the Y pixels, with two flanking axes at $\sim 67.7 \degree$ to either side.
The long wavelength arrays have a more pronounced crosshatch than the short wavelength arrays.
The NIRCam B5 array's primary crosshatch direction is most closely aligned with a pixel axis (the Y direction in the Fourier domain) of any SCA.
Figure~\ref{fig:crosshatchModelGallery} shows the best-fit crosshatch 2D power spectra.

\subsection{Position of SUBGRISM Array}\label{sec:subgrismPos}

The crosshatch pattern varies in amplitude and angular dependence from one physical location on the detector to another.
The A5 detector has more pronounced crosshatching toward the middle of the detector that falls off toward the perimeter.
On the other hand, the B5 detector has more pronounced crosshatching towards the boundary.
We fit the crosshatch pattern to three regions of the A5 detector, corresponding to those used for the NIRCam grism time series mode as well as another considered position:
\begin{enumerate}
	\item The 2040$\times$64 SUBGRISM position at the bottom of the array where the bottom left corner is (4,5) \label{subgrism64pos}, which excludes an extra illuminated row (see below)
	\item A 2040$\times$64 theoretical subarray at the top of the A5 detector where the bottom left corner is (4,1984)
	\item A 2040$\times$64 cutout of the full frame detector that is centered on the grism time series field point where the bottom left corner is (4, 249).
\end{enumerate}
These three regions are depicted in Appendix \ref{sec:regionMap}.
\added{
The SUBGRISM256 and SUBGRISM128 modes will place the target at the same location as the 2048$\times$64 SUBGRISM64 position, so results would apply to all 3 subarrays.}

\begin{figure}[!hbtp]
\centering
\includegraphics[width=0.99\columnwidth]{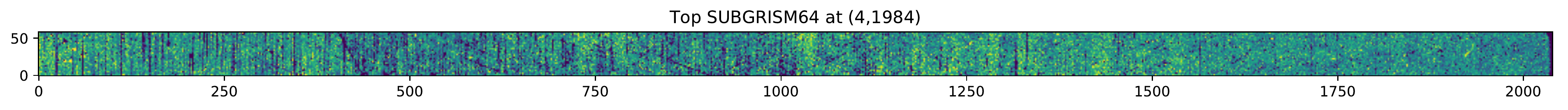}\\
\includegraphics[width=0.99\columnwidth]{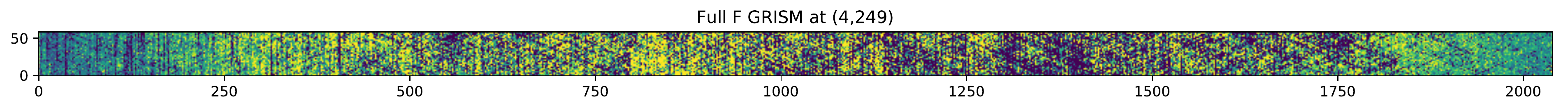}\\
\includegraphics[width=0.99\columnwidth]{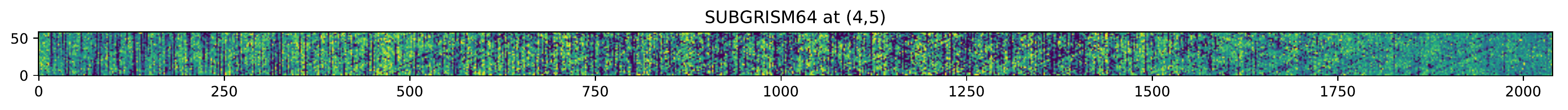}\\
\vspace{0.2in}
\includegraphics[width=0.99\columnwidth]{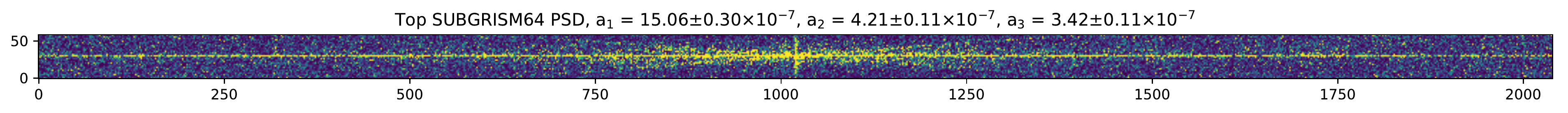}\\
\includegraphics[width=0.99\columnwidth]{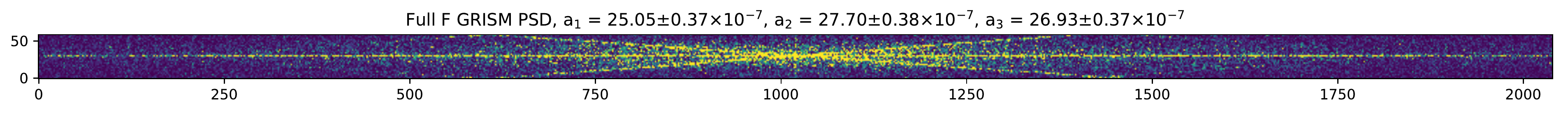}\\
\includegraphics[width=0.99\columnwidth]{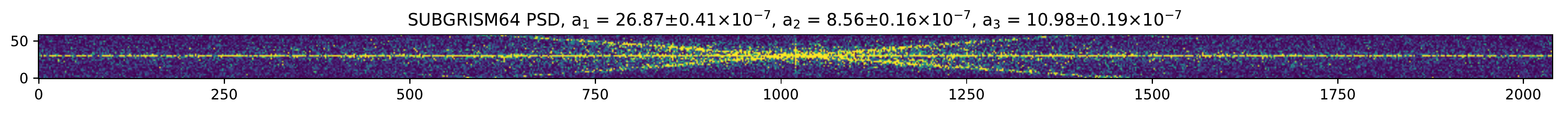}\\
\caption{
The top of the A5 array has the smallest level of crosshatching and the middle has the largest.
This is visible in the flat field images (top three panels) and their power spectral density functions (bottom three panels) for three different sections of the A5 detector: a theoretical ``aperture'' position at the top of the array (first plot from top), the full frame position near Y=512 (second plot from top) and the nominal SUBGRISM64 at the bottom of the array (third plot from top).
The bottom left corner for each subarray is listed in parentheses in the title of the flat field plot and the best-fit amplitudes are listed in the title of the power spectral density plot.
 }\label{fig:crosshatchGrismPos}
\end{figure}

The three regions considered are shown in Figure~\ref{fig:crosshatchGrismPos}.
The coordinates of these regions' corners are specified in raw detector pixels.
We exclude reference pixels (which form a 4 pixel wide boundary around the detector) and an additional row on the bottom and top of the array because these illuminated rows are outliers in the flat field.
The first position is nearly the same as the SUBGRISM64 subarray position in the grism time series mode, except that the SUBGRISM64 includes the first 5 rows which are excluded here from the crosshatch analysis.
The field point used for the SUBGRISM64 mode is expected to be the same one as will be used for the SUBGRISM128 and SUBGRISM256 subarrays.
This field point is X=468, Y=35 for the F322W2 filter and X=1097, Y=35 for the F444W filter.
The position at the top of the A5 detector has the smallest amplitude of crosshatch pattern in all three directions.
The best-fit amplitudes are used as input to the jitter simulations in Section~\ref{sec:CrosshatchSim}.
We find that the maximum flux change for a 0.1 pixel shift is 118 ppm for the nominal grism position, 108 ppm for the top of the array and 153 ppm at the full frame grism position.

\subsection{Subpixel Crosshatching Calculation}\label{sec:CrosshatchSim}

\begin{figure}[!hbtp]
\centering
\includegraphics[width=.32\columnwidth]{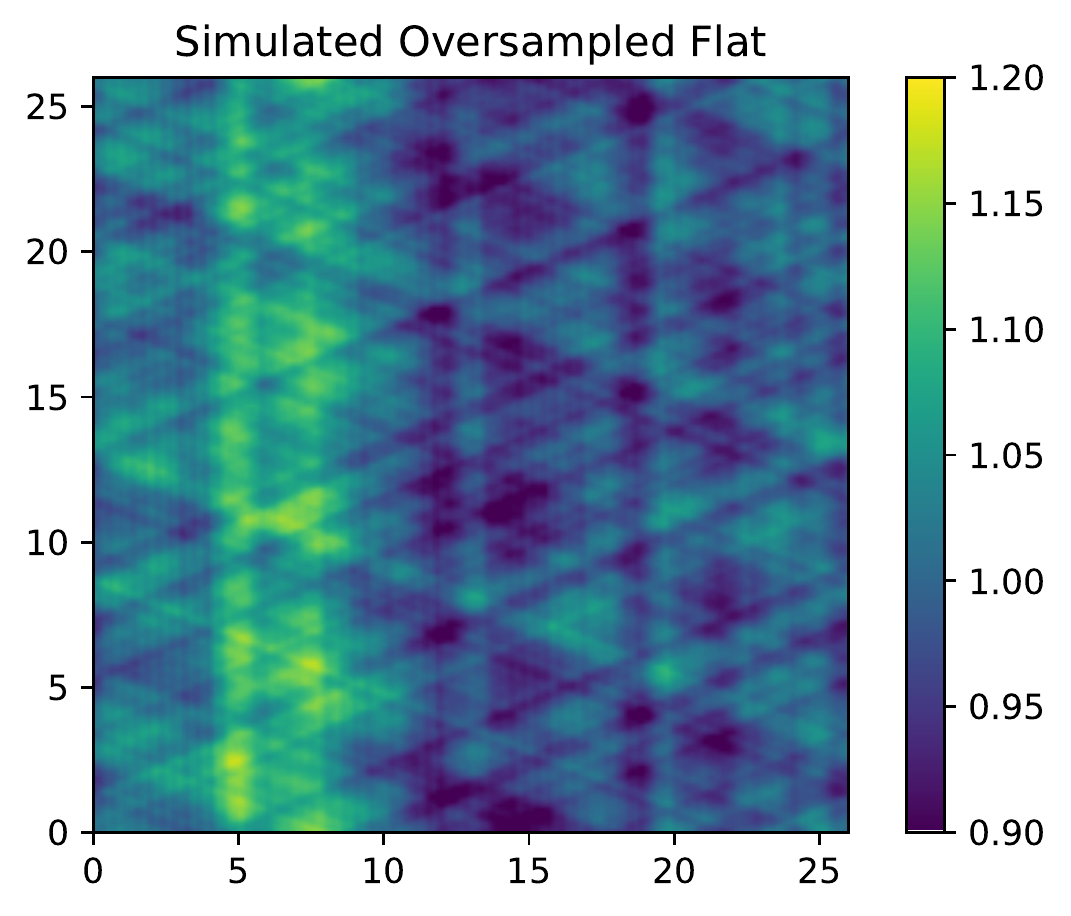}
\includegraphics[width=.32\columnwidth]{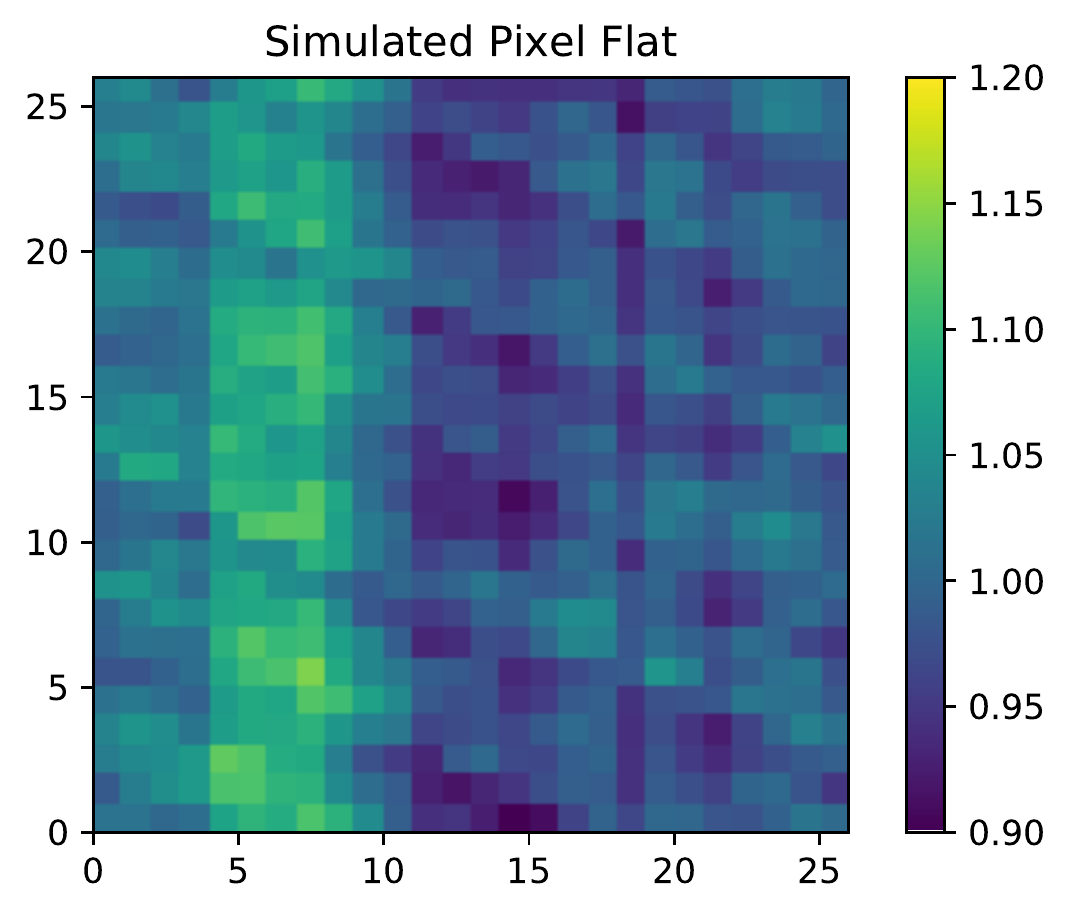}
\includegraphics[width=.32\columnwidth]{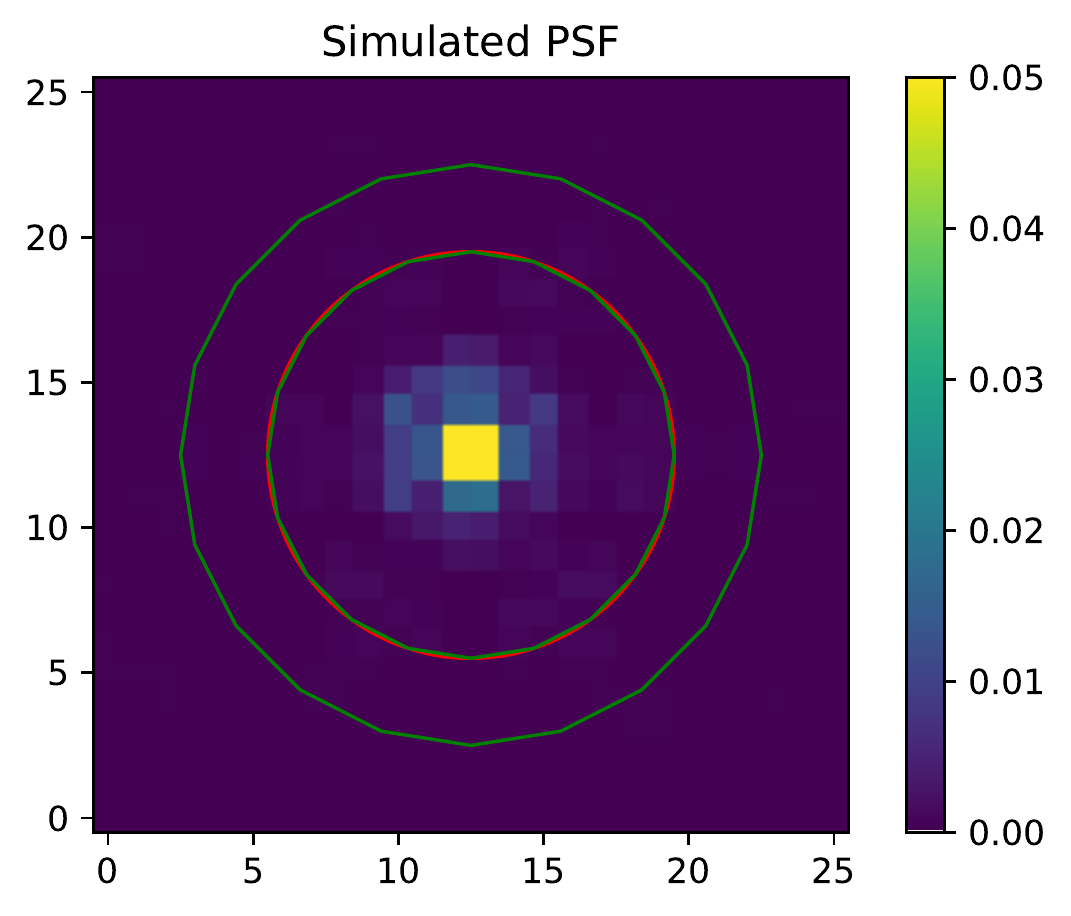}
\caption{
The steps of the sub-pixel crosshatch photometry simulation are shown here for the A5 detector.
We simulate an over-sampled flat (left) by extrapolating the power spectrum to frequencies above the pixel sampling level.
This simulated flat is binned to the native pixel resolution to produce a pixel flat field (middle) that would be used by a standard pipeline that has no sub-pixel corrections.
A \texttt{WebbPSF} point spread function is created at the oversampled resolution, multiplied by the oversampled flat field and then binned to native resolution (right).
Finally, the simulated image is divided by the pixel flat field as would be performed in a standard pipeline reduction.
The simulated PSF shows the source and background extraction apertures used in the simulation, which are centered on the PSF.
}\label{fig:crossHatchSimPSF}
\end{figure}

The subpixel crosshatch pattern can introduce time-variable noise as the point spread function moves with telescope pointing jitter and long timescale drifts.
NIRCam observations are expected to have a root-mean-square deviation of 6.0 mas in each axis when measured in 15 second intervals over a 10,000 second observation.\footnote{See \url{https://jwst-docs.stsci.edu/jwst-observatory-hardware/jwst-pointing-performance\#JWSTPointingPerformance-Pointing_stabilityPointingstability}}

We simulate the effect of this jitter on NIRCam time series observations with the following steps:
\begin{enumerate}
	\item Fit the existing flat field to a crosshatch model in the Fourier domain for a given filter
	\item Extrapolate the crosshatch model to higher frequencies (30 px$^{-1}$) to estimate the sub-pixel structure
	\item Simulate an over-sampled PSF using \texttt{webbpsf} \citep{perrin2014webbpsf}
	\item Create a simulated image and its photometric flux \label{it:simImg}
	\begin{enumerate}
	\item Shift the PSF along the X or Y direction \label{it:xHatchShiftPSF}
	\item Multiply the PSF by the simulated flat field
	\item Bin the simulated images into the native pixel size
	\item Divide by the pixel-to-pixel flat field (as would be done in a pipeline)
	\item Measure the flux within photometric extraction aperture with the same centroid shift applied \label{it:xHatchApPhot}
	\end{enumerate}
	\item{Repeat Step \ref{it:simImg} for a scan along the X direction across +/- 6 mas}
	\item{Repeat Step \ref{it:simImg} for a scan along the Y direction across +/- 6 mas}
\end{enumerate}

As discussed, in Section~\ref{sec:crosshatchModeling}, we fit Equation~\ref{eq:fullPSDmodel} to the measured power spectral density of the F300M filter for the frequencies above 0.05 px$^{-1}$.
We evaluate Equation~\ref{eq:analyticPSDradial} for frequencies in the oversampled image from 0 px$^{-1}$ to 30 px$^{-1}$ and multiply this by the number of pixels and the square of the oversampling factor (i.e. $60^2$) to convert from the scale-invariant Power Spectral Density to the simulated PSD.
We then assign random phases uniformly from 0 to 2$\pi$ for the complex Fourier plane because the original image's complex phase distribution is similar to uniform.
Finally, we take the real part of the inverse Fourier transform to create an oversampled image, as shown in Figure~\ref{fig:crossHatchSimPSF}.
For comparison to the original flat in Figure~\ref{fig:crossHatchA5Zoom}, we binned the oversampled flat field to the native LW pixel (1.0 native pixels=60.0 oversampled pixels $\approx$0.063\arcsec).
This simulated pixel flat field with dimensions of 26$\times$26 LW pixels has a peak of 1.15 to valley of 0.90 with a robust (outlier-rejected) standard deviation of 0.05.
For comparison, the middle of the original F300M flat for the A5 detector has a robust standard deviation 0.06.

We next calculate a point spread function using \texttt{webbpsf} \citep{perrin2014webbpsf}, oversampled by a factor of 60 for the F300M filter as a representative wavelength for the F322W2 filter.
\texttt{webbpsf} includes an optional blurring due to high frequency pointing jitter.
We set this Gaussian blurring parameter to 1~mas to simulate a worst-case scenario where the pointing drift is dominated by long-timescale behavior with minimal high frequency jitter.
This is a worst-case scenario because the sharpest images will result in the most sub-pixel sensitivity.
We multiply the oversampled \texttt{webbpsf} source by the simulated oversampled flat and bin this to native pixel resolution to create a simulated observation as shown in Figure \ref{fig:crossHatchSimPSF}.
This simulated observation is divided by the binned simulated flat as would be done by a standard pipeline's flat field correction.

We repeat steps \ref{it:xHatchShiftPSF} through \ref{it:xHatchApPhot} of the observation simulation by shifting the over-sampled PSF by sub-pixel amounts.
The subpixel shifts of the over-sampled PSF are performed with \texttt{scipy.ndimage.shift}.
For each shifted PSF, we multiply this by the subpixel crosshatch pattern and bin the result.
These simulated operations represent pointing drift on the subpixel scale.

\begin{figure}[!hbtp]
\centering
\includegraphics[width=.39\columnwidth]{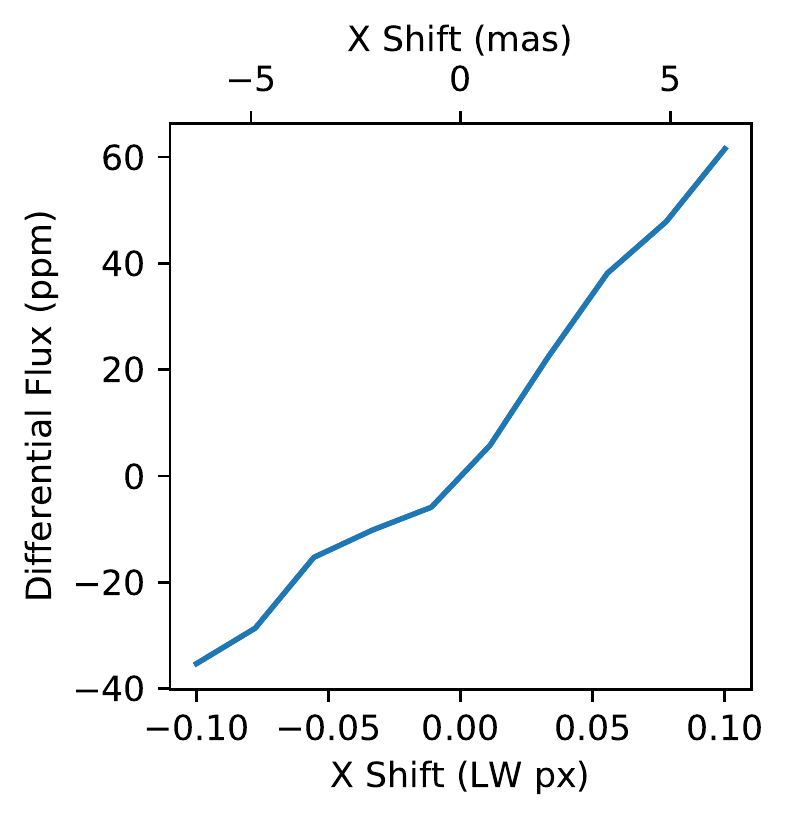}
\includegraphics[width=.39\columnwidth]{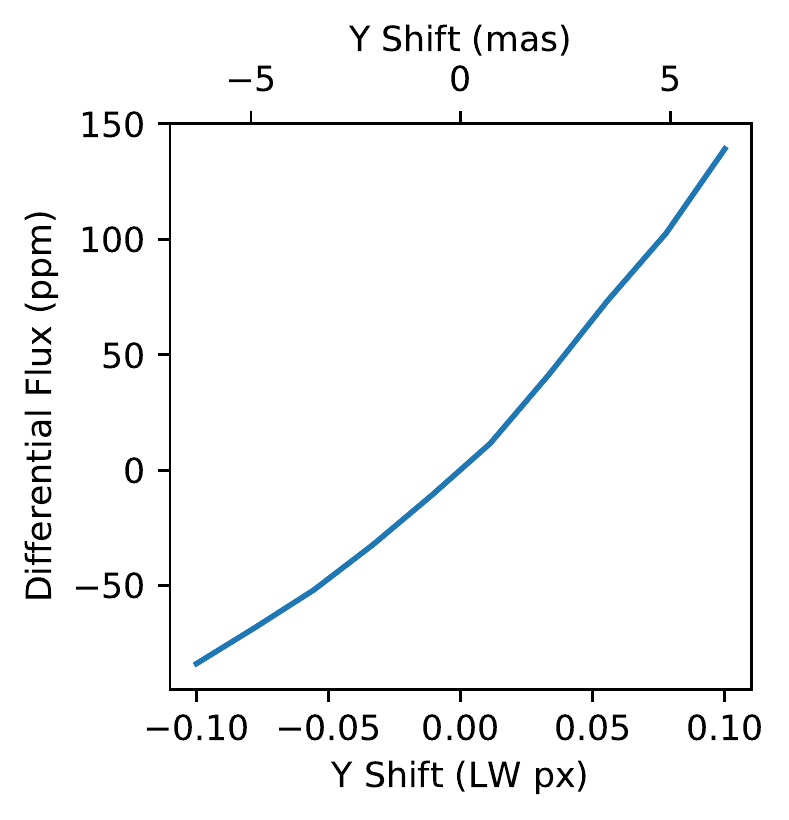}
\caption{
The simulated subpixel crosshatch pattern from Figure~\ref{fig:crossHatchSimPSF} for the F300M filter and the A5 detector will cause spurious flux changes with small pointing drifts.
Here, we show scans in the X and Y directions of $\pm$6.3~mas for the $\sim$63~mas/pix plate scale of the LW detector.
These are correctable with smooth functions such as polynomials.
}\label{fig:subpixScanSimulation}
\end{figure}

We calculate aperture photometry with a circular extraction aperture using a radius of 7 pixels and a background annulus from 7 to 10 pixels on each of the observations.
The aperture is re-centered by the same position as the shift direction.
For each sub-pixel pointing drift, the aperture sum is subtracted by the background flux average per pixel multiplied by the pixel area of the source extraction aperture.
While the simulations here include no background flux, we use the background subtraction to simulate the operations applied by a photometry pipeline.
These are also expected to be similar to a spectroscopic pipeline for a narrow spectral region or spectral line.
The differential flux between the centered PSF and the shifted one is shown in Figure~\ref{fig:subpixScanSimulation}.

\subsection{Subpixel Crosshatching Scan Results}\label{sec:subpxCrossHatchScanResults}

The subpixel crosshatch structure does indeed create flux variations with image motion as shown in Figure~\ref{fig:subpixScanSimulation}.
The amplitude of the flux changes is potentially up to 150 ppm.
This is significant when compared to atmospheric features of giant planets ($\lesssim$ 100 ppm) or the transit depth of an Earth-like planet transiting a sun-like star (84 ppm).

Fortunately, the subpixel crosshatch systematic is a smooth function of pointing drift, shown in Figure~\ref{fig:subpixScanSimulation}.
A polynomial function can be fit to this subpixel dependence and then the centroid of each integration can be inserted as an argument to the function to provide a correction as a function of time.
Alternatively, a Gaussian process regression or Pixel Level De-correlation \citep[PLD][]{deming2015secE} could be applied.
Centroiding is possible in imaging and spectroscopic modes using PSF fitting or cross-correlation.
When the long-wavelength (LW) grism time series mode is enabled, short wavelength (SW) imaging data will automatically be collected simultaneously, enabling the SW centroids to be used to track the motion of the dispersed grism image.
The SW centroids should be better than $\sim$1 mas using a Gaussian fit to the a central spot in a weak lens image.
Furthermore, the pixels for time series modes can be characterized in detail because they will be re-used for all time series observations in a given mode.
Target acquisition is expected to achieve centroiding accuracy $\lesssim$10~mas ($\lesssim 0.15$ long wavelength pixels) for unsaturated target acquisition.\footnote{See https://jwst-docs.stsci.edu/near-infrared-camera/nircam-operations/nircam-target-acquisition/nircam-time-series-imaging-target-acquisition}
This ensures that time series observations will reliably return to the same location within the same set of pixels with every visit.

\begin{figure}[!hbtp]
\centering
\includegraphics[width=.49\columnwidth]{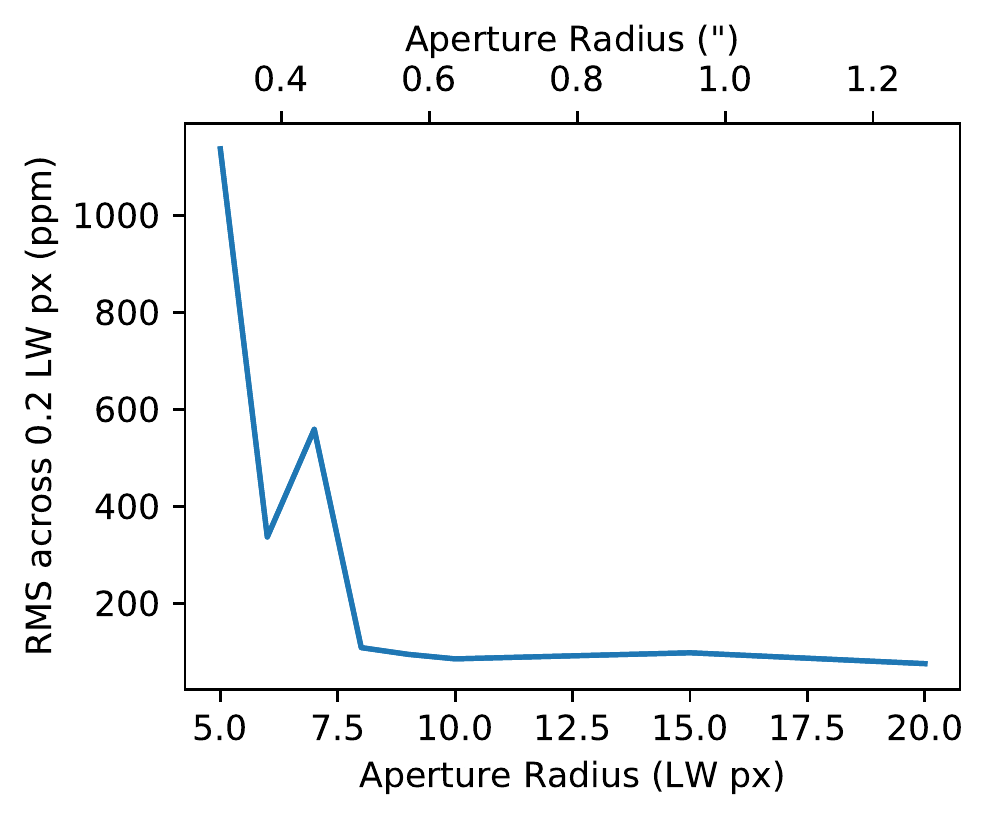}
\caption{
It is important to use sufficiently large ($\gtrsim 8$ LW pixels) extraction apertures, because the flux variability due to jitter grows significantly for smaller radii.
Here, we simulate a -6.3~mas to +6.3~mas scan in the Y direction in 10 steps as shown in Figure \ref{fig:subpixScanSimulation} (Right) and measure the standard deviation of the differential flux.
The experiment is repeated for a range of source extraction aperture radii, where a clear threshold in behavior is visible at about 8 LW pixels.
}\label{fig:subpixRMSapRadius}
\end{figure}

We find that the flux variations due to the crosshatch pattern are very sensitive to the source extraction aperture radius.
As in Figure \ref{fig:subpixScanSimulation} (right), we calculate a scan of sub-pixel shifts along the Y direction from -6.3~mas to +6.3~mas (for a a total length of 0.2 LW pixels at the 63~mas/px plate scale).
Next, we calculate the standard deviation of flux across this scan to characterize its variability.
We repeat this calculation for source apertures from 5 to 20~LW~px to explore the variability as a function of aperture size.
As shown in Figure \ref{fig:subpixRMSapRadius}, there is a clear threshold of about 8~LW~px or 0.5\arcsec, below which the flux variability grows strongly.
This is likely due to the exponential nature of the crosshatch pattern's power spectrum described in Equation \ref{eq:analyticPSDradial}.
Therefore, it is important to extract an extraction aperture greater than 8 LW px to reduce the sensitivity of NIRCam photometry or spectroscopy to the detectors' crosshatch patterns.
This ensures a lower amplitude of variability with pointing jitter as well as a smoother function of centroid position.
We note that this is in tension with the issue of 1/f noise discussed in \paperI\ because the read noise favors small apertures or strong weights on the brightest pixels.

\subsection{Time Series Simulation}

\begin{figure*}
\gridline{\fig{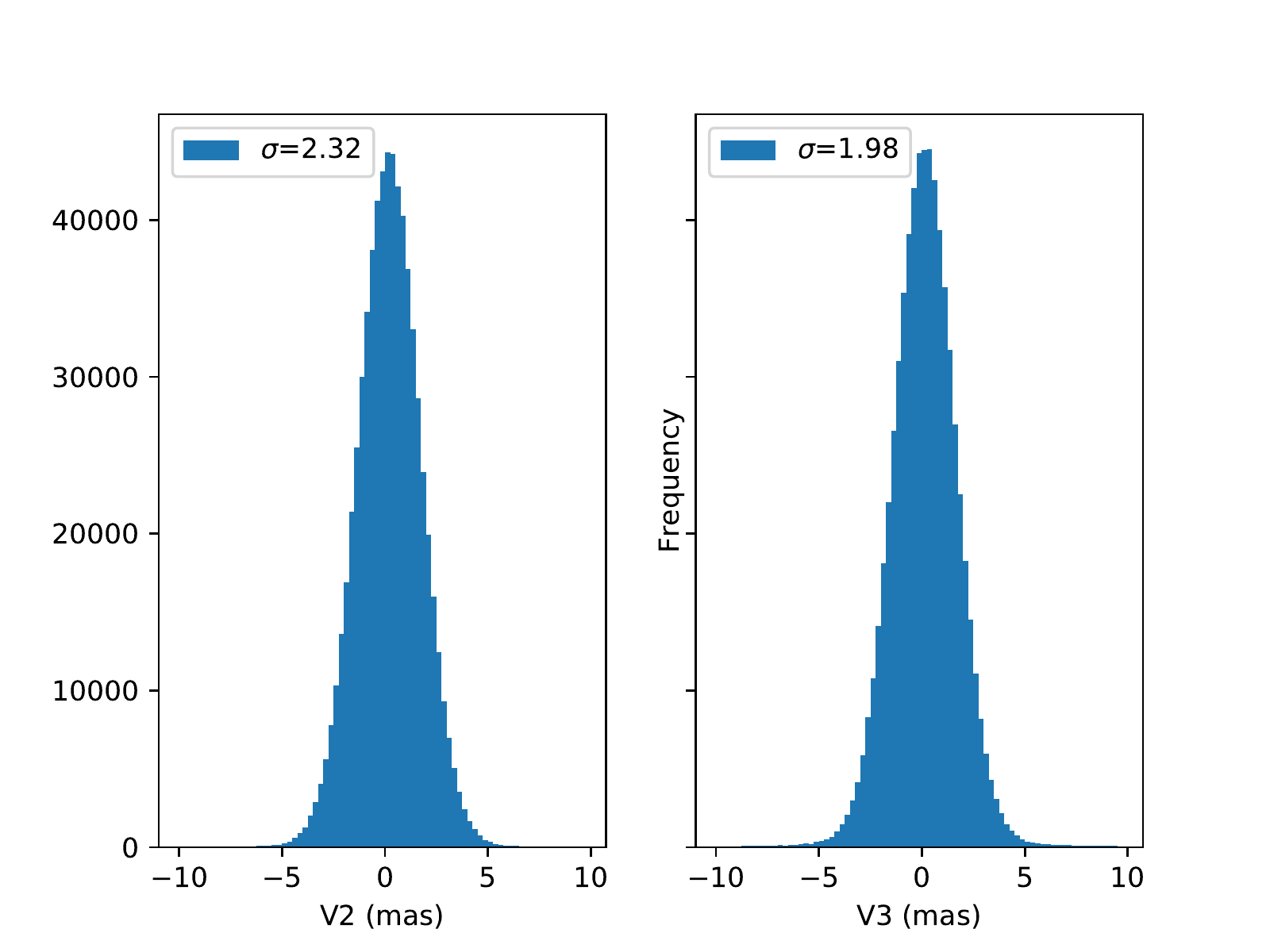}{0.45\textwidth}{Pointing Histograms} 
\fig{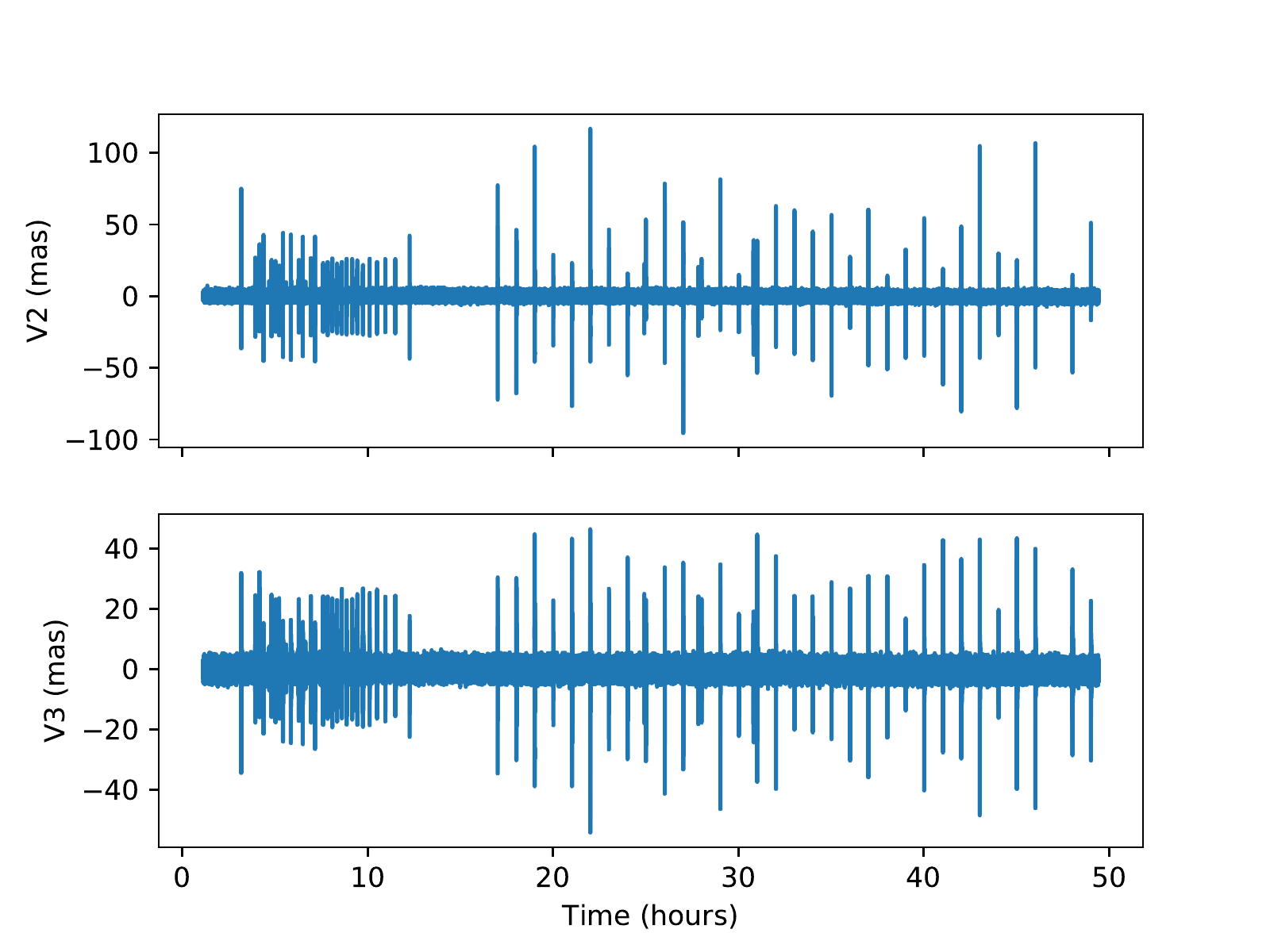}{0.45\textwidth}{Pointing Time Series} }
\caption{The typical pointing stability is expected to be extremely stable to within a standard deviation of 2.4~mas (0.04 long wavelength pixels), as seen in the histograms (left).
There are occasional jumps due to High Gain Antenna repointings (right).
This pointing model only shows the line of sight variations, not including thermal distortions or the high frequency ($>$1 Hz) jitter.
\deleted{The} V2 and V3 are the observatory axes perpendicular to the telescope boresight axis.
}\label{fig:tserPointingFull}
\end{figure*}

\begin{figure*}
\gridline{\fig{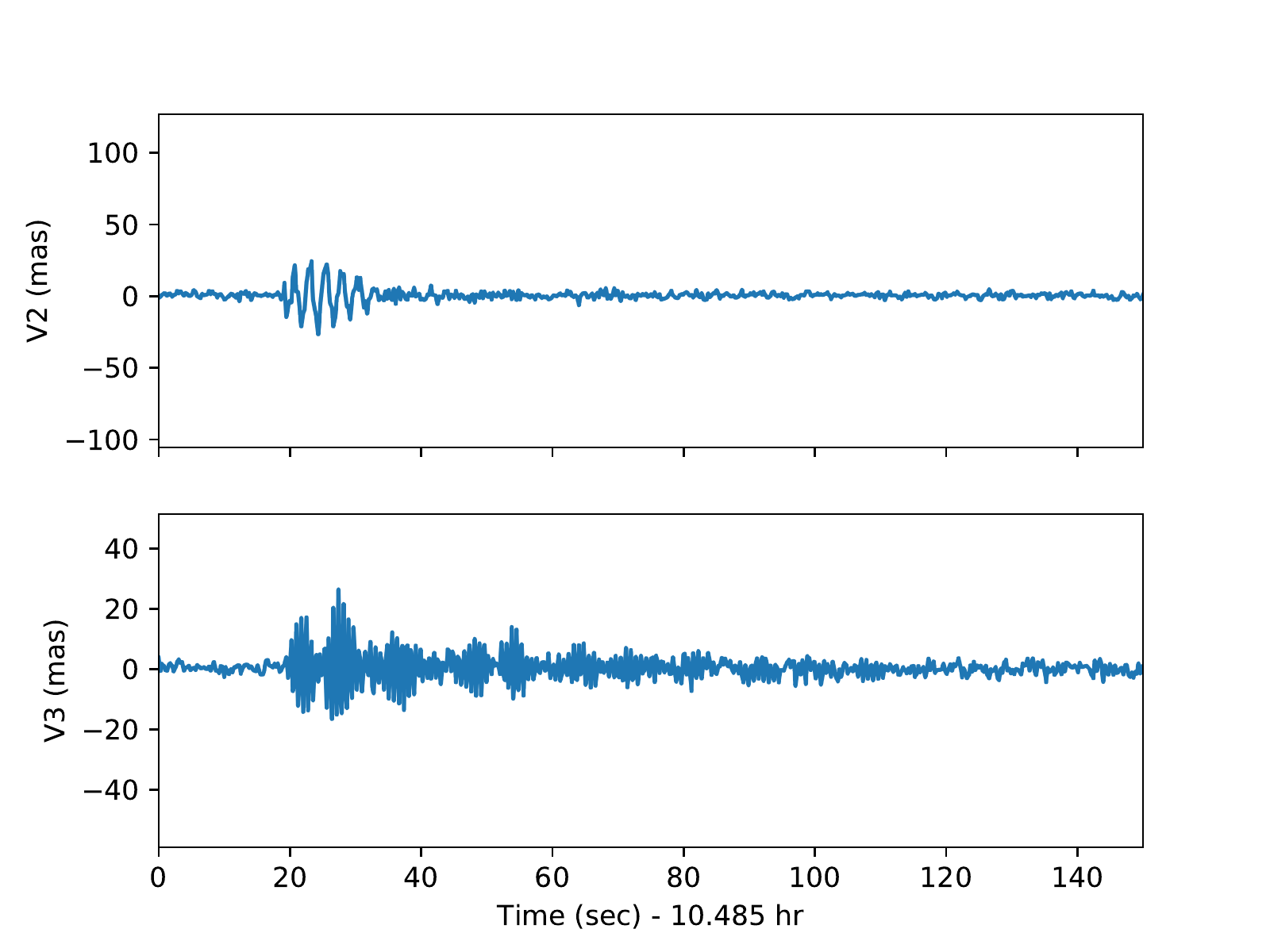}{0.45\textwidth}{Pointing Series Zoom 1} 
\fig{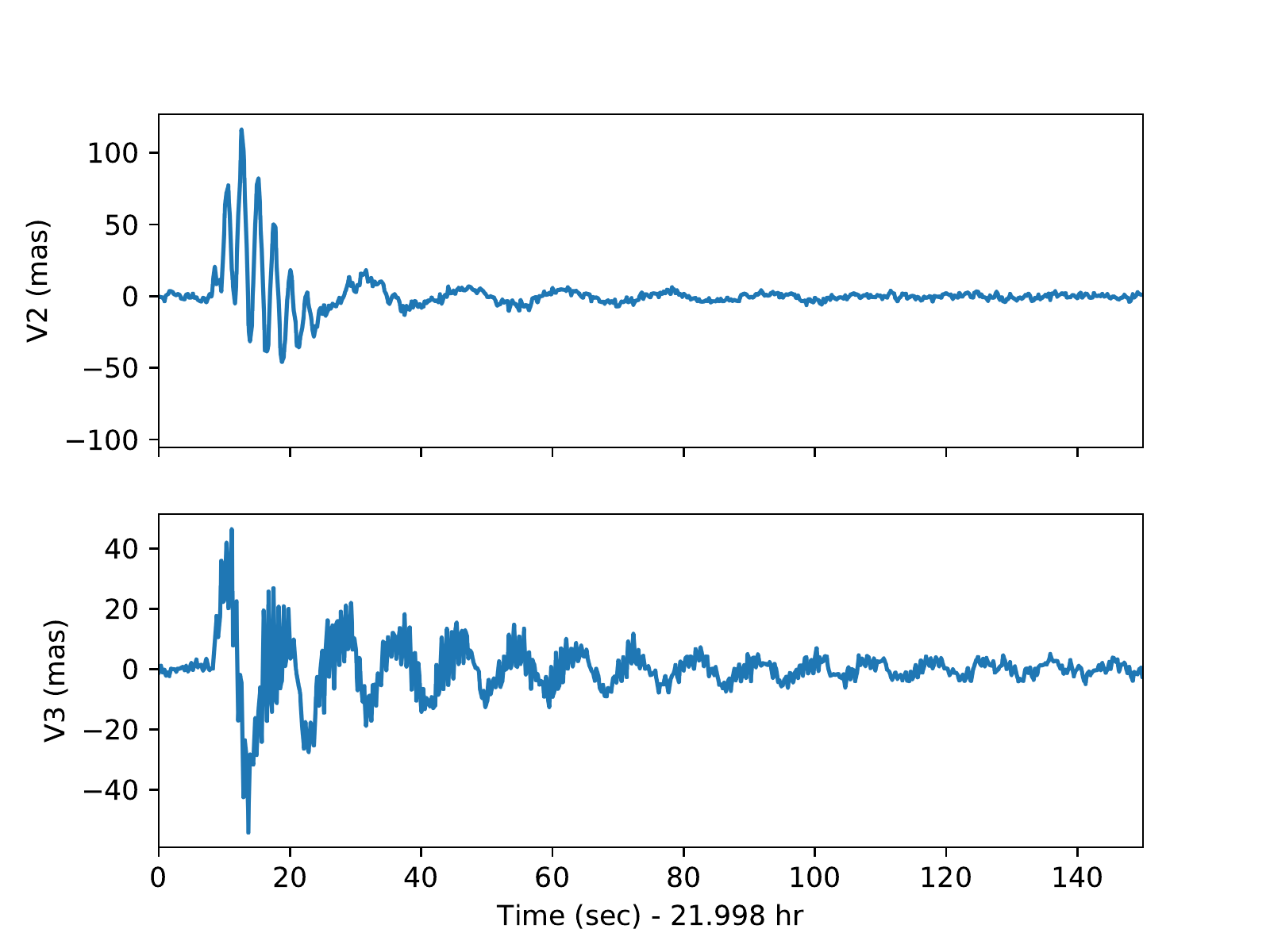}{0.45\textwidth}{Pointing Series Zoom 2} }
\caption{Occasional pointing errors due to High Gain Antenna repointings are expected to be quickly corrected by the fine steering mirror with less than 1 minute e-folding timescales.
}\label{fig:tserPointingZoom}
\end{figure*}

\begin{figure*}
\gridline{\fig{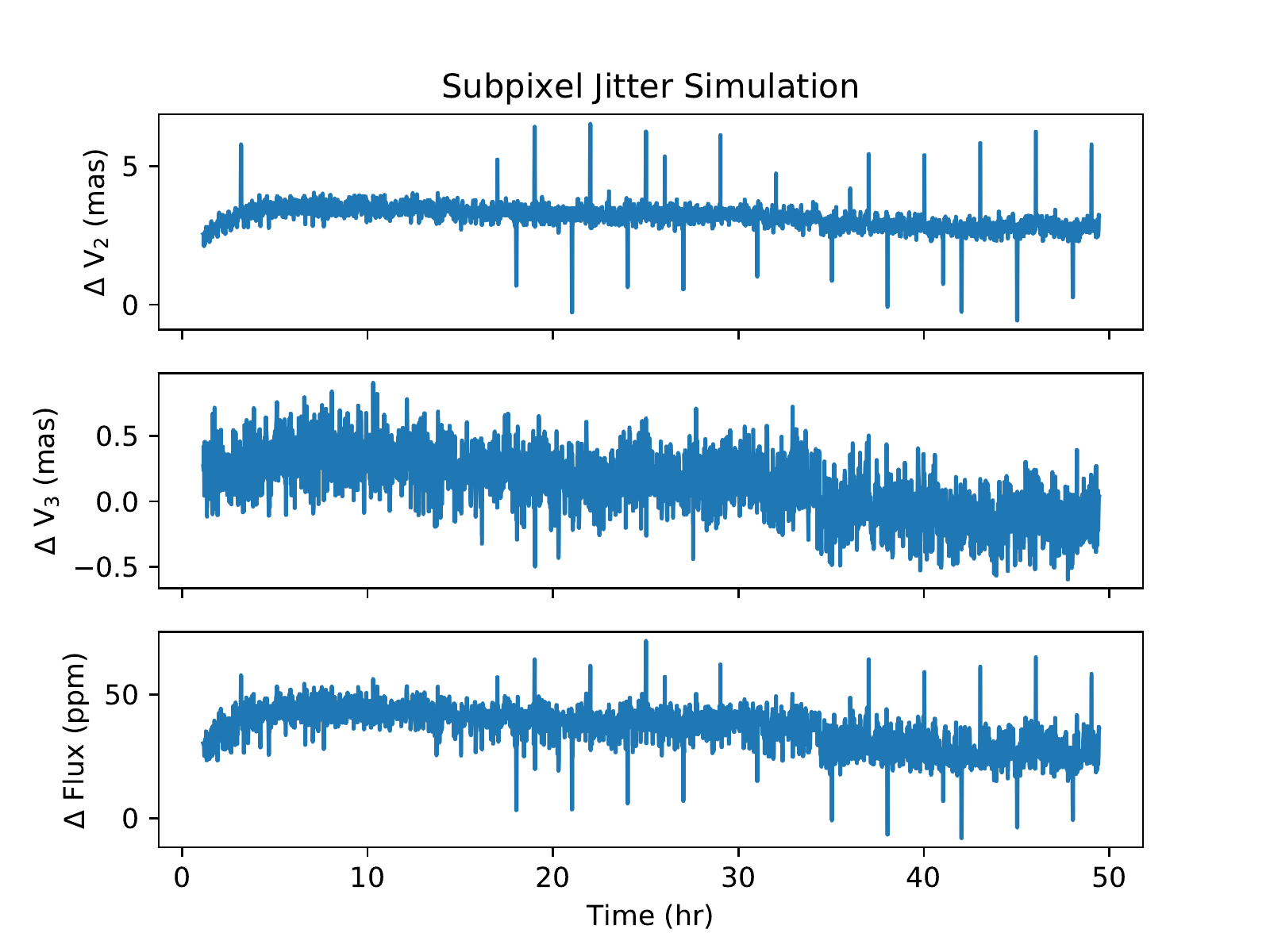}{0.7\textwidth}{}}
\caption{The pointing time series from Figure~\ref{fig:tserPointingFull} is binned into a 1 minute cadence and a thermal term is added using Equation \ref{eq:thermalPointing}, resulting in a 1-minute cadence time series (top two panels) and used as input shifts to the crosshatch model.
The very precise pointing expected by JWST will produce very small flux variations (bottom panel) on the crosshatch pattern with an expected standard deviation of only 6 ppm.
}\label{fig:tserPointingAndFlux}
\end{figure*}

\deleted{The JWST line of sight pointing contributors can be categorized into slow ($<$1 Hz) and fast ($>$1 Hz) disturbances.  The JWST Attitude Control System (ACS) is designed to correct the slow disturbances in closed loop using the fine guidance sensor (FGS) camera as the sensor and the fine steering mirror as the corrector.  However, these slow disturbances are sensed and corrected in the V2 and V3 planes (`X,Y' in detector coordinates), and there is an uncorrected term that includes a roll about the telescope boresight.  The slow drifts include disturbances from fuel slosh or thermal distortion between the star trackers and the telescope.  The fast disturbances are due to exported vibrations from the reaction wheels, MIRI cryocooler, science instrument mechanisms, and high gain antenna maneuvers that are mostly above the fine guidance control bandwidth.  The JWST Attitude Control System team provided a simulation of the expected pointing performance based on ground-test data, which is shown in Figure \ref{fig:tserPointingFull}.}\added{The JWST Attitude Control System team provided a simulation of the expected pointing performance based on ground-test data, which is shown in Figure \ref{fig:tserPointingFull}.
The pointing is tracked and corrected by the the Fine Guidance Sensor (FGS) instrument but there remain small pointing residuals and drifts on-sky for the NIRCam instrument.
These pointing residuals includes slow ($<$1 Hz) and fast ($>$1 Hz) disturbances that are below and above the above the fine guidance control bandwidth, respectively.
}

The slow disturbances include contributions from thermal distortion from the star trackers and the telescope, as well as fuel slosh that occurs when the telescope is re-pointed.  The thermal distortion component is affected by repointings of the telescope that change the solar illumination on the star trackers, which is mostly affected by pitch changes but also include effects from roll changes.  The time constant for the thermal distortion component is expect to be $\sim$ hours.   The fuel slosh is excited at the beginning of a visit when slewing to the target and stopping at that new attitude.  For both of these disturbances, there is an impact on the roll about the guide star that is not corrected by the low bandwidth coarse pointing loop.
\deleted{In other words, the pointing is tracked and corrected by the the FGS instrument but there remain small pointing residuals and drifts on-sky for the NIRCam instrument.}

The fast disturbances come from the reaction wheels, the pulsed MIRI cryocooler, science instrument mechanisms, and high gain antenna (HGA).
\added{These fast disturbances occur at rates $>$1 Hz that are outside the FGS bandwidth}.
The reaction wheel jitter is time-variable due to changes in wheel speeds as momentum accumulates, whereas the MIRI cryocooler jitter is pulsed at a constant frequency that is tuned to avoid resonances with the Observatory's deployed dynamics.  The reaction wheels and MIRI cryocooler are passively attenuated but will still lead to a blurring of the PSF over an integration. The science instrument mechanism and high gain antenna disturbances are intermittent only and managed by the operational procedures. In the case of transiting exoplanet observations, it is preferred to stay under fine guidance control to keep the star on the same pixels even when mechanisms or the HGA is moved.  When the high gain antenna is adjusted, there are larger disturbances, as shown in Figure~\ref{fig:tserPointingFull}.  These HGA moves are designed to keep the antenna pointing at ground-based antennas for telemetry and data download.  The errors due to HGA moves are predicted to dampen quickly on $\lesssim$ 1 minute e-folding timescales, as show in Figure \ref{fig:tserPointingZoom}.
This means that for 1 minute integration times, only about 1\% of integrations will be affected by HGA moves and the data for these moves could be discarded.

\deleted{In addition to the line-of-sight disturbances discussed, there is inter-boresight motion where the science instrument detectors move due to the thermal expansion and contraction of the telescope.
This inter-boresight motion on the science instruments is not correctable with the star trackers nor fine guidance sensor since they are not co-located with the science instruments.
The prediction from observatory models (P. Maghami, private communication) is that the inter-boresight motion is spatially linear in the V2/V3 plane and has a time constant that is about $\sim$1 hour.
We model the thermal disturbance with an exponential function,}
\added{In addition to the line-of-sight disturbances discussed, there is inter-boresight motion due to the thermal distortion of the star trackers and the telescope. 
The inter-boresight motion results in a centroid shift of sources on the science instruments.
While the V2/V3 plane is sensed and corrected by the fine guidance control loop, there is a small residual uncompensated roll about the V1 axis due to telescope and star tracker distortions.
The prediction from observatory models is that the uncorrected inter-boresight motion can be approximated as spatially linear in the V2/V3 plane and has a time constant that is ~1 hour (P. Maghami, private communication). 
We therefore model the disturbance with an exponential function,}
\begin{equation}
V_{thermal} = 2.55 mas \left(1 - e^{-t/\tau} \right),
\label{eq:thermalPointing}
\end{equation}

where $V_{thermal}$ is the thermal drift that is in a random linear direction in the V2/V3 plane, the magnitude of the shift is 2.55 mas and $\tau$ is the exponential constant that is set to be 3600 sec \added{(P. Maghami, private communication)}.
\added{In reality, the time constant for the star tracker assembly and optical telescope element are different, but this represents a worst case for the slow drift.}
This thermal term is then added to the above line of sight jitter simulation to produce an estimate of the pointing variations for the NIRCam instrument.

\added{The 2.55 mas thermal drift amplitude in Equation \ref{eq:thermalPointing} was derived using the worst case lever arm from a guide star to any point in the field of view of NIRCam' A-side ($\sim$10.06\arcmin) and can vary across the focal plane.
The boresight term is mainly due to the observatory's roll, and is proportional to the magnitude of the vector from guide star and science target star.
For example, a variation in the field point of 1\arcmin\ along the same line would result in a reduction of image motion of around 2.55 mas $\times$ 1\arcmin/10.06\arcmin=0.25 mas.
The magnitude of pointing errors for MIRI, NIRSpec and NIRISS are expected to be a factor of 1.6, 1.1 and 1.2 larger, respectively (P. Maghami, private communication).
}

We simulate the change in flux as a function of position using our oversampled flat field, as in Figure~\ref{fig:subpixScanSimulation}.
First, we bin the sub-pixel jitter time series into 1 minute time bins to represent 1 minute integrations, as shown in Figure~\ref{fig:tserPointingAndFlux}.
We then use the positions to shift the PSF on the oversampled flat field as listed at the beginning of this section.
Next, we perform photometry on the simulated images with a circular aperture and background annulus.

The resulting flux variations \added{on each integration} are very small compared to factors like the 1/f noise.
The standard deviation of the time series due to small line-of-sight telescope jitter on top of the subpixel crosshatching pattern is only 6 ppm.
Therefore, the \replaced{line-of-sight subpixel}{pointing} jitter is expected to have a small affect on NIRCam time series stability.
Furthermore, centroid motion measurements can be used to correct the flux variations due to subpixel jitter with a polynomial or Gaussian process regression.
\added{Small aperture sizes that are most affected by subpixel motions (below 0.5\arcsec\ in radius) would benefit the most from a centroid correction function.}
Another possibility for modeling the crosshatch pattern is a fit to the Fourier amplitude and phase as done for the flat field but in the sub-pixel regime.
The spacecraft telemetry can be used to identify High Gain Antenna moves to assess the short ($\lesssim$ 1 minute e-folding) centroid vibrations it causes and remove that data if desired.

We note that the 6 ppm value will be wavelength and position dependent, so it is just a representative value at 3.0\micron.
This flux variation can vary with both the wavelength-dependence of the PSF and the location of a given wavelength on the detector on the spatially-dependent crosshatch pattern.
For example, we showed in Section \ref{sec:subgrismPos} that at 3 different field points, the flux variations for a 0.1 pixel shift varied from 108 ppm to 153 ppm.
Therefore, the noise due to pointing jitter of 6 ppm is expected to vary by about 50\%, so it can range from 3 ppm to 9 ppm depending on the particular wavelength.

\section{Variable Aperture Losses}\label{sec:varApertureLosses}

\begin{figure*}
\gridline{\fig{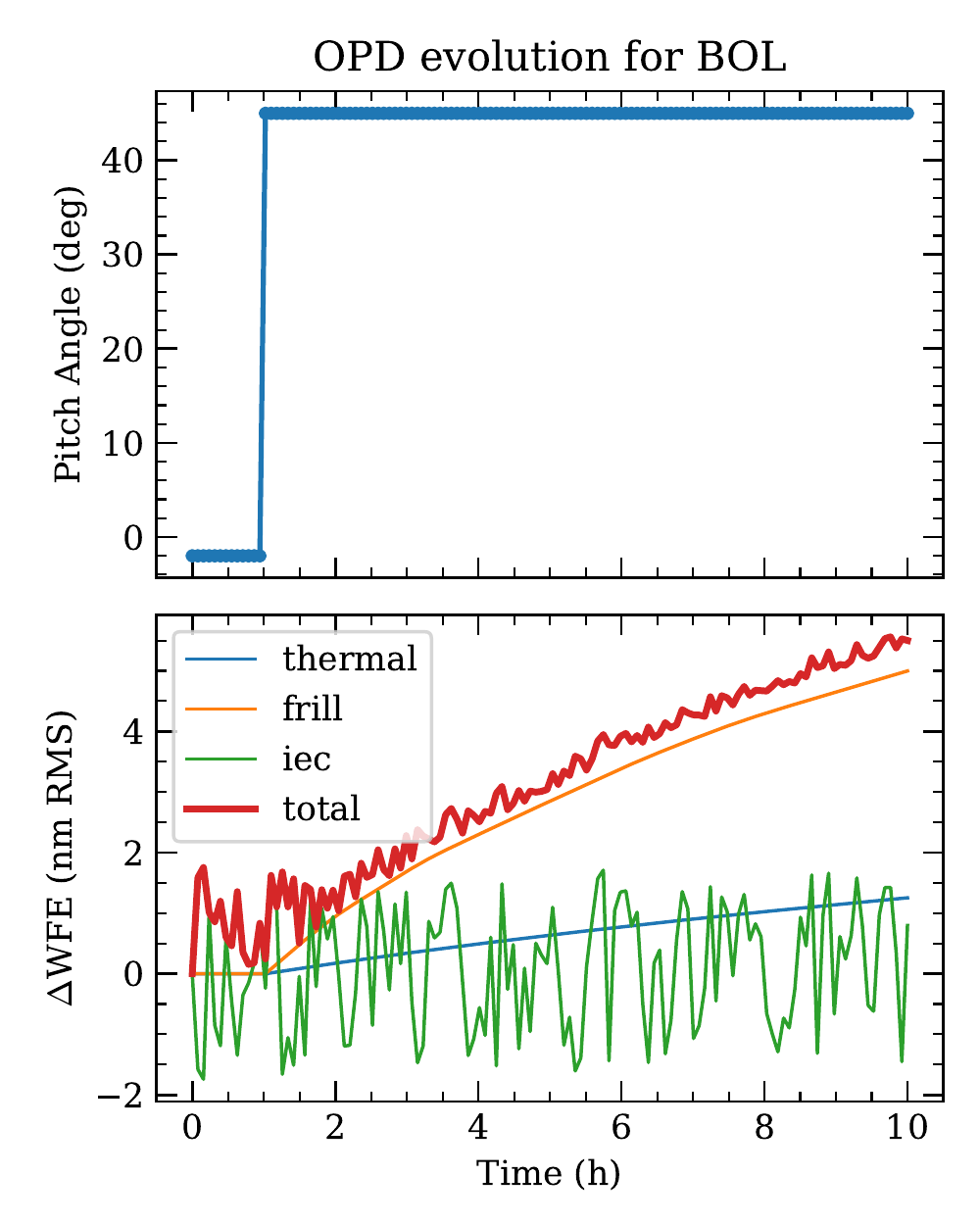}{0.45\textwidth}{OPD evolution}}
\caption{The wavefront error (WFE) evolves in time due to thermal changes of the observatory.
Here, we assume a single slew to the exoplanet science target (top panel), which causes slow telescope and frill changes (blue and orange lines). 
On top of these variations, the IEC heaters (green line) cause short timescale variations in the wavefront. Values assume performance for beginning of life (BOL).
}\label{fig:opdEvolution}
\end{figure*}

JWST was designed to keep the optical performance near the diffraction limit and minimize disturbances to the wavefront.
However, there are small optical path differences (OPDs) from an ideal wavefront that can change with time.
The mirrors can undergo slight deviations in position, shape and angle due to the thermal expansion and contraction of the observatory components.
Further, there can be oscillations or vibrations of the observatory components in response to mechanical disturbances such as reaction wheels and the cryocooler.
These variations in the OPD will result in different PSF images within the NIRCam instrument.
A changing PSF can adversely affect time series observations by varying the amount of flux contained within a fixed extraction aperture, much like it does for ground-based observing at faster and more dramatic timescales with atmospheric seeing variations.

We assume JWST begins by observing an unrelated object for a different science program at a  ``hot`` attitude or pitch angle of -2$\deg$ and then it is slewed to an exoplanet system to a ``cold`` attitude with a significantly different pitch of 45$\deg$.
This large slew would represent the worst case for thermal differences in the time series.
Figure \ref{fig:opdEvolution} shows the time evolution of the OPD wavefront error (WFE) in response to this slew.

We use a parameterized version of a Northrop Grumman and NASA Goddard thermal models incorporated into \texttt{webbpsf}.
These models account for changes in the primary mirror \replaced{Zernicke}{Zernike} components from the three main \replaced{components}{contributions} shown in Figure \ref{fig:opdEvolution}: 1) temperature perturbations that occur on the Optical Telescope Element (OTE) backplane as the telescope pitch angle changes during a slew (blue), 2) tensioning of the primary mirror frill structure (orange), 3) and effects of the ISIM Electronics Compartment (IEC) heaters (green).
The total effect of all these WFE sources is shown in red.
This model is based on measurements from cryogenic vacuum testing at NASA Johnson where the telescope was cooled from $\sim$300 K to $\sim$40 K.
The time constant for the thermal changes is about 5-6 days, so a typical exoplanet transit or eclipse observation ($\sim$hours) will not reach thermal equilibrium.
We assume beginning of life (BOL) conditions where the sunshield has had no degradation due to micrometeroid impacts.

\begin{figure*}
\gridline{\fig{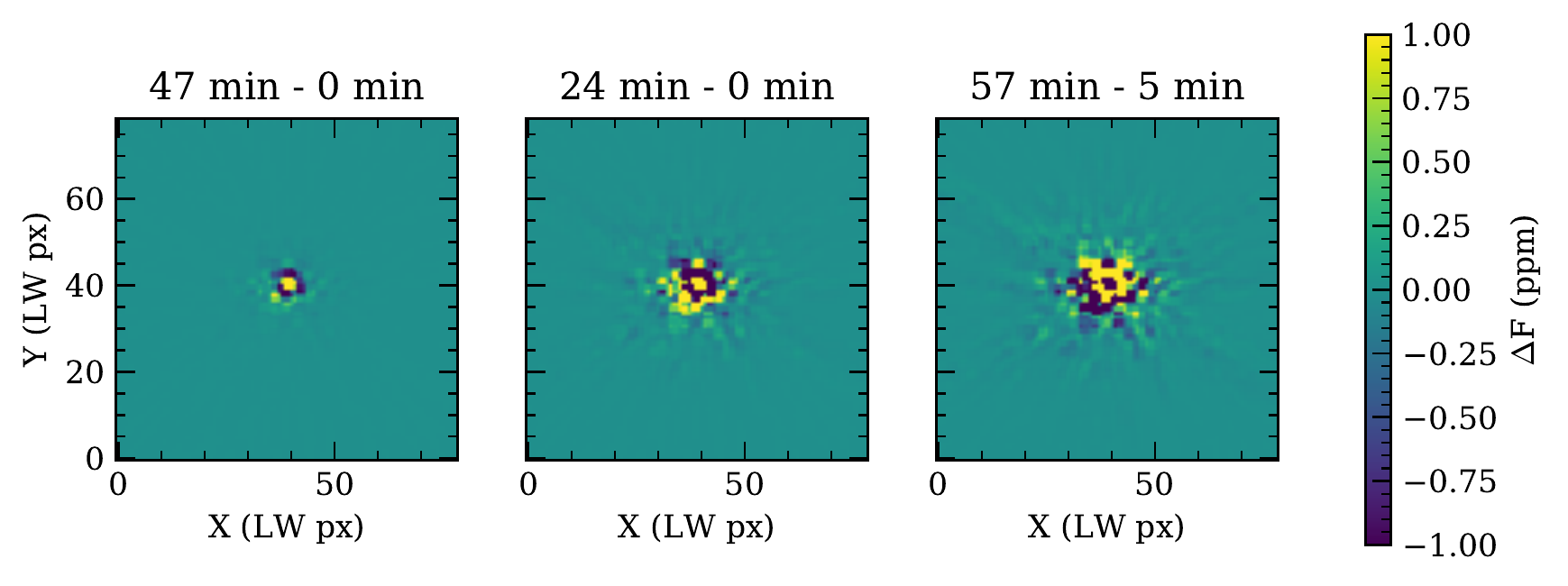}{0.45\textwidth}{IEC heater PSF differences}\\
\fig{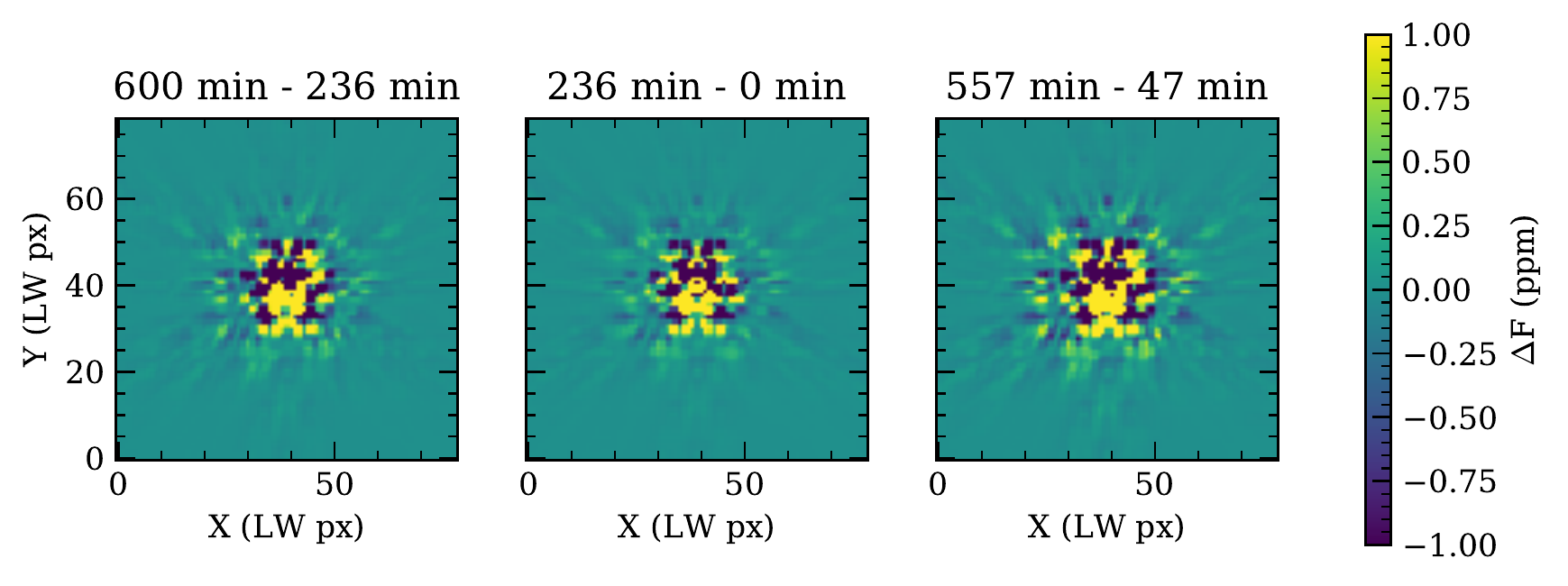}{0.45\textwidth}{Thermal PSF differences}}
\caption{Wavefront errors will change the point spread function (PSF) with time, shown here for the F444W filter.	
PSF differences are largets for the core ($\lesssim$5 LW px) due to the integrated science instrument module electronics compartment heater (left) and largest at intermediate distances ($\sim10$ LW px) due to thermal variations (right).
\added{The PSF differences shown here are selected from specific times in the entire simulation to highlight the morphology and variation that are possible when the PSF changes.}
}\label{fig:psfEvolution}
\end{figure*}

We calculate the PSFs for each simulated OPD using \texttt{webbpsf} \citep{perrin2014webbpsf} to examine how this can impact time series.
Figure \ref{fig:psfEvolution} shows the differences in the PSFs at different epochs of the time evolution.
In the beginning, we assume the primary mirror and frill are in thermal equilibrium so the only wavefront disturbances come from the IEC heaters, which change the core of the PSF ($\lesssim$5 LW px or $\lesssim$0.32\arcsec), as shown in Figure \ref{fig:psfEvolution}, left.
After 1 hour in our simulation, the telescope is slewed and the thermal and frill changes become important.
Now, the PSF differences are most pronounced at intermediate distances ($\sim$10 LW px or 0.63\arcsec).

\begin{figure*}
\gridline{\fig{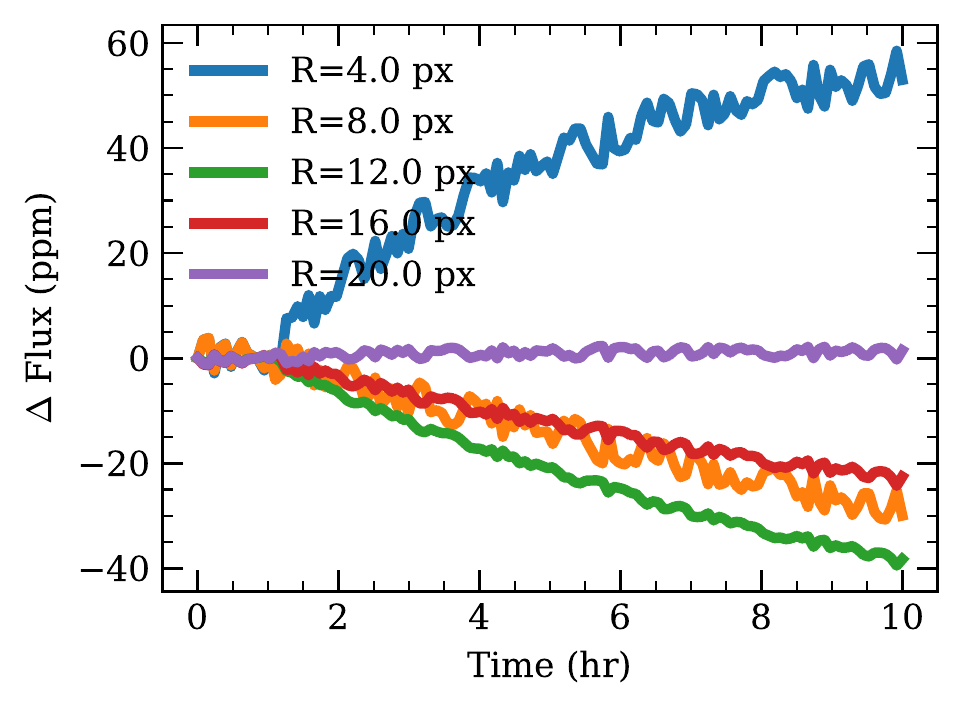}{0.45\textwidth}{} 
\fig{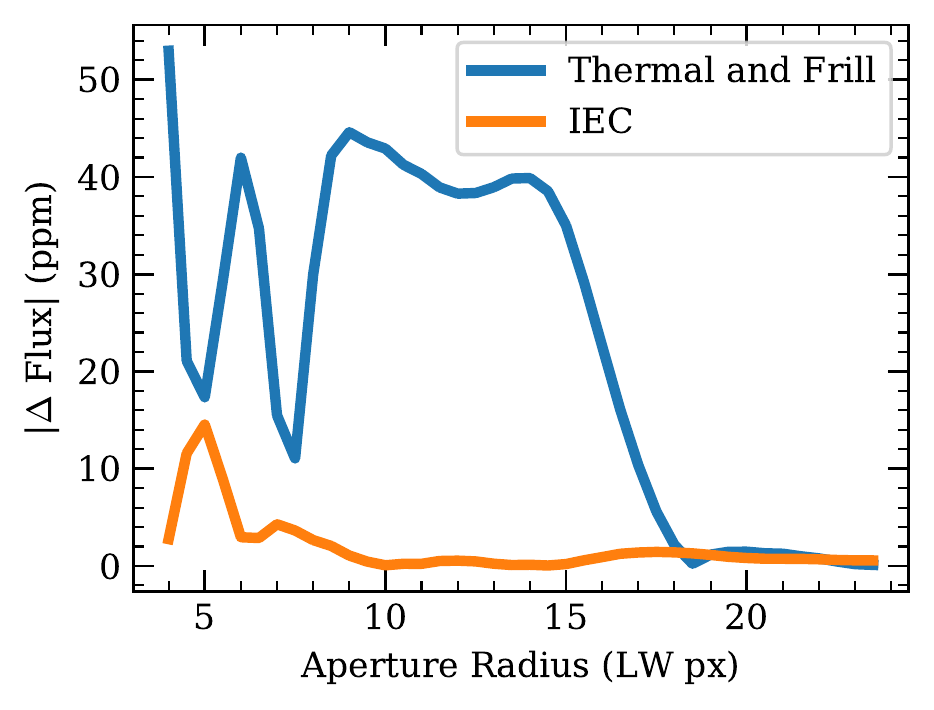}{0.45\textwidth}{} }
\caption{Wavefront errors will cause flux losses within an extraction aperture, causing variations in time series.
{\it Left:} The flux will vary smoothly along the 5-6 day telescope thermal timescale, but small apertures are sensitive to the IEC heaters.
{\it Right:} Small apertures are more sensitive to WFE, but above 18 pixels, the time variations are less than 2 ppm.
}\label{fig:apPhotWFE}
\end{figure*}

We next examined how the PSF variations manifest as time-variable extraction aperture losses.
We calculated aperture photometry using the \texttt{photutils} package and experimented with a range in aperture radii from 4 to 23 LW px (0.25\arcsec\ to 1.5\arcsec).
We centered the apertures by fitting a 2 dimensional Gaussian.
The resulting changes in flux are shown in Figure \ref{fig:apPhotWFE} in parts per million.

In the variable-wavefront time series, the largest variations are from slow thermal variations of the wavefront.
From the start of the slew at 1.0 hours to the end of the simulation at 10.0 hours, the variations can be as much as 55 ppm.
The thermal variations cause smooth changes in the aperture flux, so they can be removed with linear or quadratic de-trending.
This will likely be smaller than the host star flux variability.
For example, the typical solar variability on minutes-long to hour-long timescales is 50 ppm or larger, as measured by the Total Irradiance Monitor \citep{kopp2005tmi}.
The stellar variability tends to grow larger with later spectral types \citep{rackham2019lightSourceEffect2}.

We find that the variable aperture losses also \replaced{grow}{are} smaller for large photometric apertures.
For small extraction apertures ($\lesssim$6 LW px or $\lesssim$0.4\arcsec), there is also a 5-10 ppm variability due to the IEC heaters, as seen in Figure \ref{fig:apPhotWFE} because the IEC heaters affect the core of the PSF.
The thermal variations affect the PSF farther out, so an aperture radius larger than 10 LW px (0.63\arcsec) is needed to minimize the thermal disturbances.
With a sufficiently large apertures ($>18$px), the time variable aperture losses go down to 2 ppm.
We separated the IEC component and thermal components in Figure \ref{fig:apPhotWFE} by choosing images dominated by those components.
For the IEC-dominated component, we compare the fluxes for image 6 and 1 (before the slew or thermal changes happen) and for the Thermal-dominated component, we compare the last and first images.
We note that the aperture loss experiments assume a circular fixed aperture with uniform pixel weights, but another \replaced{covariance}{weighting} scheme might be applied such as the covariance weights in \paperI.

The variable aperture losses are modest ($\lesssim$ 60 ppm) for all apertures tested here, but could be reduced with more advanced photometry.
One method commonly used for ground-based observations is to scale the aperture radius by the full width at half maximum (FWHM).
Scaling apertures with the size of the PSF has successfully reduced noise with Spitzer IRAC lightcurves \citep{knutson2012phase189,ingalls2016spitzerRepeatability}.
Here, we estimate the FWHM by fitting a Gaussian to every simulated PSF.
This does reduce the thermal PSF disturbances by $\sim$5 ppm for an aperture radius that is 7.4 times the FWHM versus an aperture radius that is fixed at 16 px.
However, an aperture scaled by the FWHM is less helpful for the fast IEC heater variations with small aperture radii.


\section{Charge Trapping}\label{sec:persistence}
\begin{figure*}
{\gridline{\fig{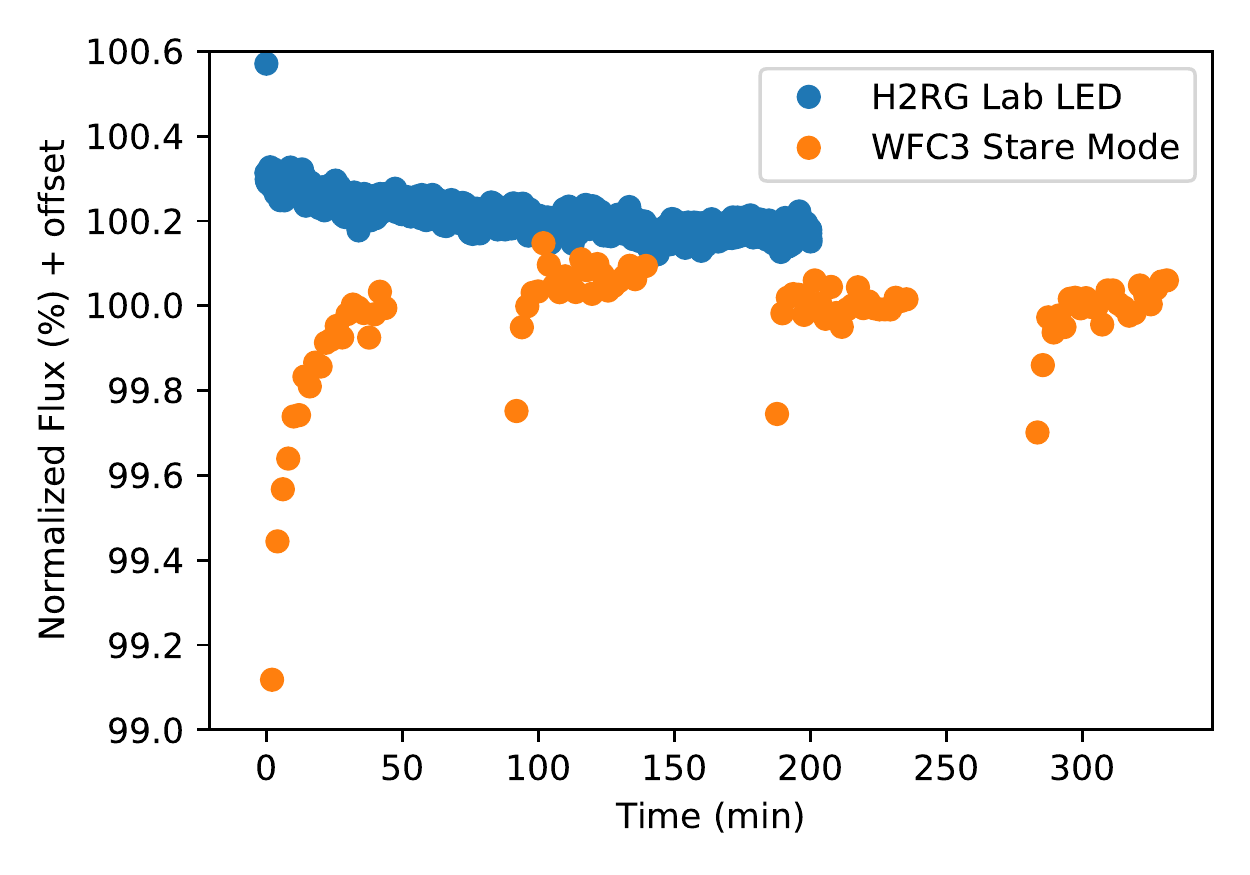}{0.5\textwidth}{}}}
\caption{The JWST H2RG detectors will have much smaller ramp amplitudes than the HST WFC3 detector.
\added{The} WFC3 lightcurve (orange points) shows a ramp-up behavior with every orbit due to charge trapping \citep[e.g.][]{zhou2017chargeTrap}.
The Earth occultations with every HST orbit prevent the detector from reaching a steady state.
JWST, on the other hand, will observe uninterrupted to approach a steady state and use H2RG detectors that have smaller charge trapping amplitudes.
A Lab-LED illuminated H2RG detector's lightcurve (blue points) shows much less pronounced charge trapping behavior.
The variations within this H2RG lightcurve are dominated by the lamp output and not the detector itself.
}\label{fig:rampWFC3vsJWST}
\end{figure*}

HST WFC3 shows noticeable ramps with each HST orbit, where the detected flux rises approximately as $1- R e^{-t/\tau}$ as function of time $t$, where R is the ramp amplitude and $\tau$ is the exponential time constant \citep[e.g.][]{berta2012flat_gj1214}, as seen in Figure \ref{fig:rampWFC3vsJWST} for an example lightcurve.
The ramp effect has been explained and modeled as the capture and release of electrons in detector charge traps \citep{zhou2017chargeTrap}.
These charge traps (also known as persistence) are responsible for latent images, where a bright exposure followed by darker illumination shows a residual image whose emission rate decays with time \citep{smith2008imgPersistence,tulloch2018persistenceH2RG,leisenring2016persistence}.
We present a schematic for the charge trapping effect in Appendix \ref{sec:chargeTrapSchematic}.

Fortunately, the JWST NIRCam charge trapping effects are much smaller than for HST WFC3.
\citet{leisenring2016persistence} find that the persistence rate after saturating the detectors and turning off the lamp is 0.5 DN/s to 15 DN/sec or 0.9$e^-$/s to 27$e^-$/s at the 30 second mark, depending on the detector.
After 1000 s, the persistence level falls to $10^{-2}~e^-$/s for the A5 detector,
the detector used for grism time series.
For a pixel that fills to about 60\% well depth in \added{a} 20 second integration, the count rate for that pixel will be 3,200 e$^-$/s.
Therefore, the persistence level compared to the signal level at 1000 s into a time series will be about 3 ppm.
This gives an order-of-magnitude estimate for the ramp effect, assuming that the charge capture timescale is not significantly longer than the charge release timescale.

Figure \ref{fig:rampWFC3vsJWST} shows a comparison between a lightcurve from HST WFC3 and a laboratory test of a NIRCam-like detector.
The HST WFC3 observation comes from secondary eclipse observations of the CoRoT-1 system (GO program 12181), taken \deleted{in} with the G141 Filter and a GRIMS128 subarray aperture.
The exposure was taken with SPARS10 MULTIACCUM mode with 16 samples for an exposure time of 100.65 seconds.
As seen in Figure \ref{fig:rampWFC3vsJWST}, there is a pronounced ramp with each HST orbit.
The first orbit has an especially large ramp compared to the others, and thus the first orbit is typically discarded in lightcurve analysis \citep[e.g.][]{evans2016wasp121H2OTiO,wakeford2017hatp26,kreidberg14}.
The first orbit, however, can be modeled with charge trapping \citep[e.g.][]{zhou2017chargeTrap}.

For the laboratory test of a NIRCam-like detector, we use a NIRCam detector that was not selected for the flight instrument, but has many of the same noise behaviors.
The detector is controlled with SIDECAR ASIC electronics, again with an unused NIRCam part in a dewar at the University of Arizona.
An exposure was commanded in RAPID read mode and a 2048$\times$64 subarray.
There were 8 groups per integration with 4 output amplifiers, resulting in an integration time of 2.72 seconds.
For this test, a 5.2~\micron\ cutoff detector is illuminated by a 3.4~\micron\ LED source with 1.5 mA of current.
This lamp current and integration setting results in a maximum pixel value of about 60\% well capacity and thus below saturation or strong non-linearities.

The exposure is started before the lamp is turned on to study the charge trapping ramp-up immediately following illumination.
The lab H2RG LED illumination test shows much smaller ramp-up behavior than HST WFC3.
In fact, there is no detectable ramp-up behavior at all because the lamp varies with a standard deviation of 370 ppm, so any ramp behavior is below that level.
Furthermore, JWST does not have to contend with Earth occultations, so the detector has the full time of the baseline (typically $>$ 1 hour) to reach a steady state.
As seen in Figure \ref{fig:rampWFC3vsJWST}, there is \added{a} gradual decrease in flux following the lamp illumination.
These effects are likely related to the LED lamp output responding to small temperature drifts rather than charge trapping itself, because the measured flux variations can change direction and magnitude with repeated experiments.
Based on the 3 ppm persistence level estimate and 370 ppm upper limit from laboratory tests, we conclude that ramp-up behaviors will play a much smaller role for JWST NIRCam as they do for HST WFC3.
A time series test during NIRCam on-orbit commissioning will help evaluate the timescale and amplitude for any settling behaviors such as charge trapping.

\section{Detector Temperature Variations}\label{sec:detectorTemps}
The NIRCam detectors will be actively controlled to an expected precision of $\lesssim$1 mK as measured in cryogenic vacuum testing of the flight hardware.
This is achieved with a focal plane heater and temperature sensor.
We estimate here how 1 mK temperature fluctuations could manifest as time-variable signals.
We note that measurements from \citet{hall2005jwstArrays} show that 100 mK variations result in 80 DN variations in the detected absolute counts.
Interpolating linearly to smaller temperature fluctuations, there will be 0.8 DN (0.8~$e^-$ for the gain assumed in \citet{hall2005jwstArrays}) per 1 mK temperature variation.
If this unevenly affected the background and source pixels in a set of time series extraction apertures, this could adversely affect time series, but only if it shows up in the slope images.

We perform another test with the same University of Arizona laboratory dewar as discussed in Section \ref{sec:persistence} to test the effect of detector temperature on exoplanet time series of illuminated slope images with background subtraction applied.
We mount two 5.2\micron\ cutoff detectors in a focal plane array (FPA) that can accommodate 4 detectors at once.
They are both illuminated by the same 3.4\micron\ LED to simultaneously monitor the flux as the FPA changes temperature.
The exposure consists of 1000 integrations of a 2048$\times$256 STRIPE subarray with 4 outputs,  8 GROUPS in RAPID readout mode (1 frame per group) and an integration time of 10.7 seconds.
This will \replaced{most}{more} directly translate to the time variability experienced in exoplanet lightcurve measurements than the absolute counts in \citet{hall2005jwstArrays}.
The detectors are illuminated by an LED while simultaneously adjusting the temperature with an FPA heater.
The detector temperature is monitored on the FPA with a Cernox sensor.
We adjust the set point so that there is a 330~mK increase in the FPA temperature and measure the time series over 200 minutes to study the variations.
We make this large temperature change and interpolate to 1 mK because there are LED lamp variations that impede our ability to measure flux changes much below 300 ppm without de-trending models and assumptions about the noise.

\begin{figure}
{\gridline{\fig{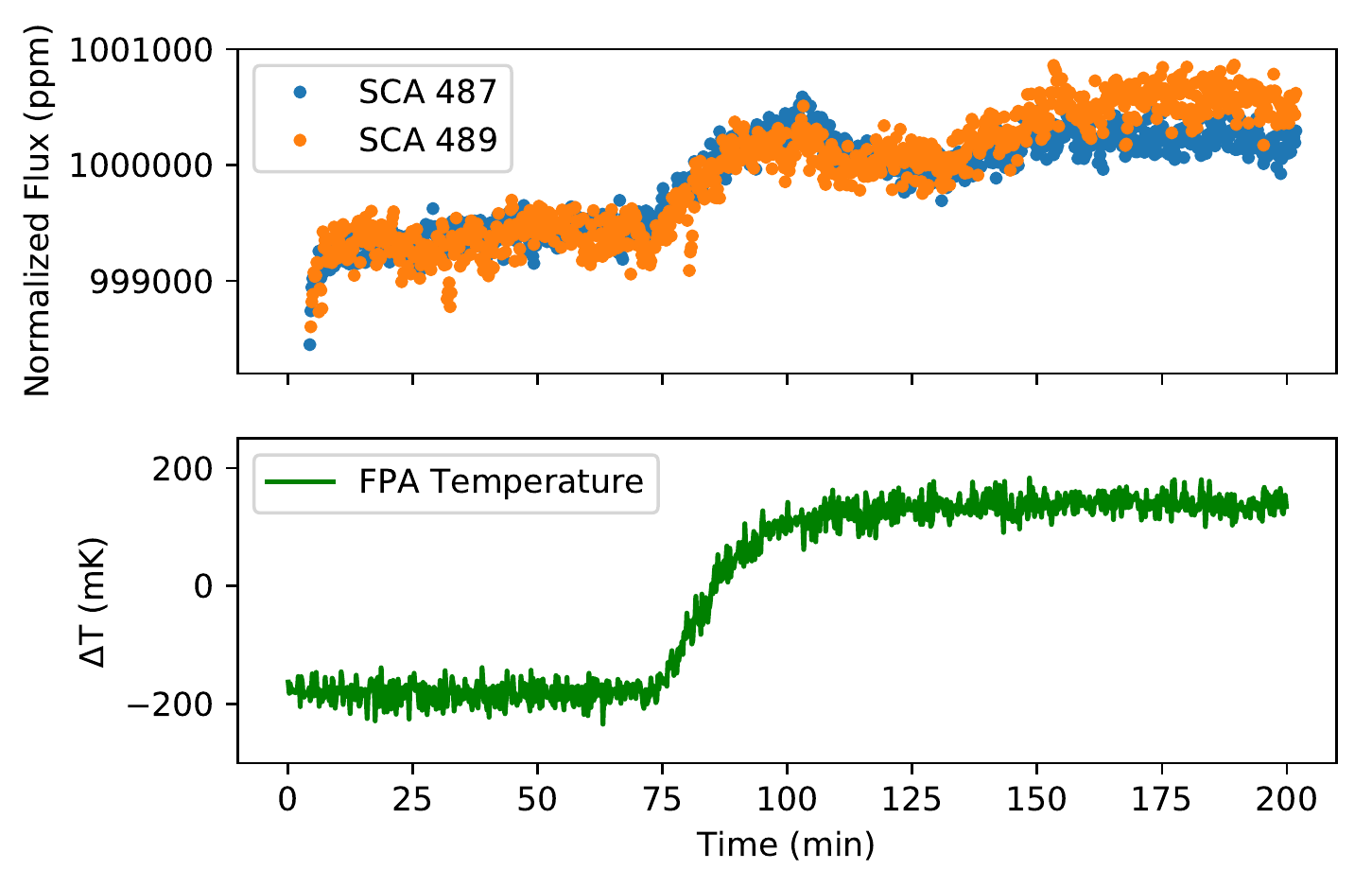}{0.5\textwidth}{}}}
\caption{The laboratory time series of two NIRCam-like detectors and electronics shows an increase in signal at higher detector temperatures.
The temperature was purposefully adjusted by \replaced{300}{330} mK here to see the effect on top of the LED Lamp variability. 
However, in flight, the temperature variations are expected to be 1 mK, resulting in only 3 ppm variations.
The dip at $\sim$130 min is most likely due to small current fluctuations in the LED.
\added{The initial 5 minute ramp is also likely due to the LED, because it was observed to change directions and magnitude depending on the test and lamp current.}
FPA telemetry can be used to calibrate temperature variations in flight.}\label{fig:fpaTempChange}
\end{figure}

Figure \ref{fig:fpaTempChange} shows the change in temperature as the heater is powered and also the resulting flux change on two different detectors.
We measure the flux with a rectangular extraction aperture of 106$\times$142 and 79$\times$180 pixels on the 487 and 489 detectors respectively, so the results represent average pixel behavior.
Both apertures' fluxes are subtracted by horizontally-offset background apertures that are 
\replaced{in shadows of the illumination pattern.}{masked off from direct illumination by the LED.}
For the 330 mK increase in temperature, both detectors (numbered 487 and 489) show approximately the same 900 ppm increase in normalized flux.
We note that the illumination levels of the two detectors are different because the peak counts per integration were 25,000 DN/px and 39,000 DN/px for detectors 487 and 489.
This suggests that the temperature variations result in fractional changes to output photometry or spectroscopy instead of an absolute change in DN.
There are variations in the lamp  output on the two detectors due to current instability or the temperature of the LED lamp itself on top of the S curve due to the temperature increase.

Next, we linearly interpolated this fractional flux increase from 330 mK variations down to 1 mK, ie a rate of 2.7 ppm per mK temperature variation.
We therefore expect detector temperature fluctuations to be a relatively small part of the time series error budget.
A caveat is that one of the temperature monitor controls (TMC1) for the A5 detector used for grism time series can result in 20 mK oscillations in \replaced{signal}{temperature}.
Fortunately, there is a redundant controller TMC2, that keeps the variations to 1 mK.
In the event that the TMC2 fails, calibrations must be applied to TMC1, which could potentially see 54 ppm variability with temperature.
Alternatively, the B5 detector could be used, albeit with a different lower transmission grism and software adjustments required to operate the subarrays for the B side of NIRCam.

\section{Reciprocity Failure and Non-Linearity}\label{sec:recipFailure}
The HgCdTe pixels are nonlinear with respect to both the percentage of their well capacity and also with the incident photon rate.
They are nonlinear with respect to the percentage of well capacity filled because the photodiode becomes less sensitive to incoming photons as the depletion region shrinks.
This non-linearity with the depletion region size (percentage of well capacity or well fraction) is calibrated and corrected in the pipeline using a constant flux source measured with different integration times.
It can also be mitigated by keeping the brightest pixels safely below the saturation level of the detector.

The nonlinearity with incident photon rate is called reciprocity failure \citep{hill2010reciprocityFailure,biesiadzinski2011reciprocityFailure} and has not been as well-characterized on the JWST detectors.
Here, we estimate the difference in flux that could occur if the NIRCam detector behaves like the HgCdTe 1.7\micron\ cutoff detector measured by \citet{hill2010reciprocityFailure}, shown in Figure 4 of that paper.
For the input signal, we assume an A0V host star, which has strong Hydrogen features and a deep transit depth of 2\%.

\begin{figure}
{\gridline{\fig{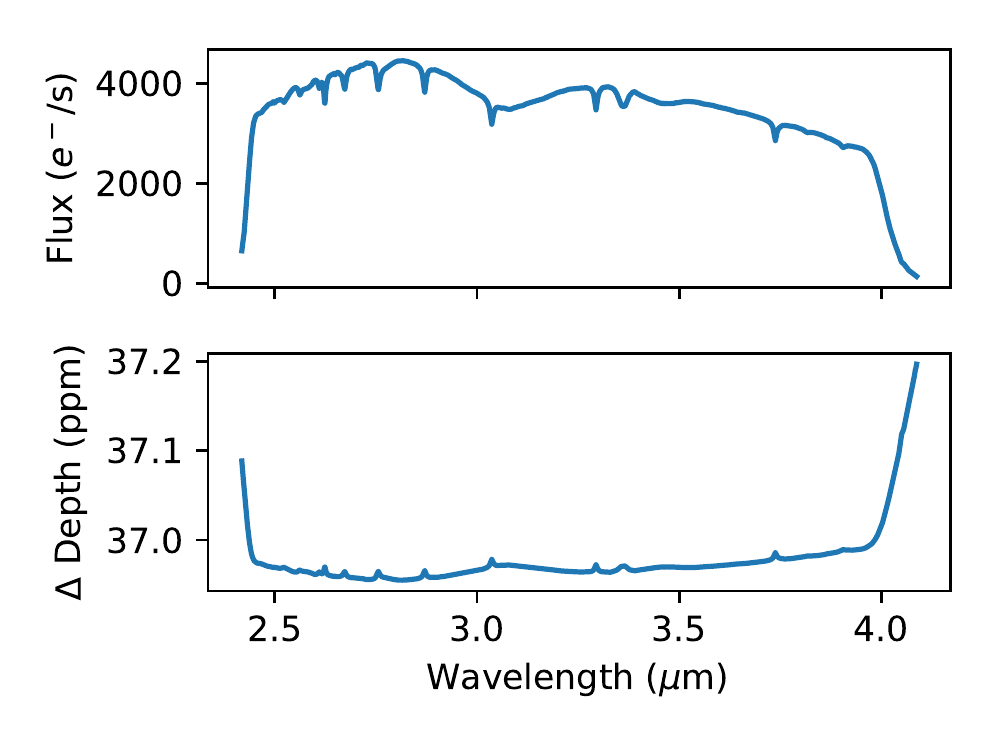}{0.5\textwidth}{}}}
\caption{Reciprocity failure will change the transit depth measured by a HgCdTe detector like the one measured in \citet{hill2010reciprocityFailure}.
The top panel shows an input A0 V spectrum signal and the bottom shows the change in transit depth caused by reciprocity failure, which adds a 37 ppm offset to the transit depth, but sub-ppm variations across the spectrum.}\label{fig:reciprocityFailure}
\end{figure}

The resulting change in transit depth due to reciprocity failure is shown in Figure \ref{fig:reciprocityFailure}.
The main effect of reciprocity failure is an offset in the transit depth.
For the 2\% (20,000 ppm) transit depth considered here, there is an offset of 37 ppm due to reciprocity failure.
The signal level changes in stellar absorption features, with the Raleigh-Jeans shape of the continuum and at the edges of the filter bandpass, but this changes the transit depth by less than 1 ppm at the wavelengths of the stellar features.
Therefore, the main effect of reciprocity failure is an offset in transit depth that could affect inter-comparison between different instruments or facilities.
We expect that residual non-linearities that are not corrected by the pipeline will have a similar effect on the signal.

\section{Conclusions and Combined Noise Floor}\label{sec:Conclusion}

JWST has the potential to transform our understanding of transiting planets with extremely high precision light curve measurements.
JWST can collect enough photons to observe atmospheric gases and even biosignatures in optimistic scenarios with no astrophysical noise \citep[e.g.][]{krissansen-totton2018trappist1eJWST}.
However, there are electronic 1/f noise effects (discussed in \paperI) and systematic effects (this paper) that can limit JWST's ability to achieve the photon limit.
We have discussed the systematic errors in this work that can appear in time series observations.

\begin{deluxetable}{ccc}


\tablecaption{Estimates of Known Systematic Errors in NIRCam Time Series}

\tablehead{\colhead{Noise Source } & \colhead{Time Variability} & \colhead{Perennial Offset} \\ 
\colhead{} & \colhead{(ppm)} & \colhead{(ppm)} } 

\startdata
Pointing Jitter on Crosshatch Pattern  & 6                         & \nodata \\
Time-Variable Aperture Losses          & 2                         & \nodata \\
Charge Traps                           & 3                         & 3 \\
Detector Temperature Fluctuations &	  3		& \nodata \\
Non-Linearity \& Reciprocity Failure    & \nodata                   & 37 \\
Ghost Images					& \nodata		& 1 \\
\hline
Total                                  & \combinedNoise \tablenotemark{*}                         & 37 \\
\enddata


\tablecomments{\tablenotemark{*}Our estimates of the known error sources indicate very small systematic levels.
We urge caution in using this as a standard because there will likely be new, unknown error sources that contribute to JWST time series errors.
We assume the pointing jitter is correlated with aperture loss but that charge traps and detector temperature are independent from all other noise terms.}\label{tab:combinedErrors}

\end{deluxetable}

We combine the known sources of systematic error here to give an estimated noise floor for the NIRCam time series modes.
Our systematic noise sources, as discussed in Section \ref{sec:subpxCrosshatch} through \ref{sec:recipFailure}, are summarized in Table \ref{tab:combinedErrors}.
We categorize the noise into two flavors that can impact the science.
The first is time variability that can occur over an exoplanet transit lightcurve or similar measurement and will likely be different from one visit to another.
The second is a ``perennial'' or permanent offset, that will appear in all visits in a similar way and could potentially affect inter-instrument comparisons when stitching a spectrum together over a wide wavelength range from the visible to near infrared.

The largest of the known \added{time-variable} noise sources is \added{due to pointing jitter on top of} the intra-pixel sensitivity.
Detector crosshatching structures extend down to the subpixel level where they will not be corrected in standard flat fielding procedures.
Fortunately, the pointing drifts are expected to be very small (2 mas) other than brief and infrequent high gain antenna moves, so the standard deviation of flux due to pointing jitter is expected to be only $\sim$ 6 ppm.
The variations in flux at the subpixel level appear smooth enough that they could likely be calibrated as a function of position.
Time-variable aperture losses play a smaller role than the subpixel crosshatching.
We estimate how heating and cooling of various observatory components can change the PSF and find that they will likely appear as 2 ppm variability.
We note that the subpixel crosshatching and time-variable aperture losses grow for apertures less than 1.1\arcsec, so wider apertures are necessary for these estimates to apply.

Charge traps and detector temperature variations will also play a small role in time-variable noise.
We estimate that charge traps will affect JWST at the 3 ppm level or smaller, which is less than HST due to low charge trap densities and because JWST will observe continuously without Earth occultations.
Detector temperatures will be controlled to 1 mK levels and thus keep the flux stable to $\lesssim 3$ ppm based on interpolation of laboratory measurements.

We combine the errors assuming that the pointing jitter and time-variable aperture losses are highly correlated and that all other errors are statistically independent.
The pointing jitter and time-variable aperture losses are both related to thermal variations and displacement of instrument optics that have similar time constants.
We therefore assume that these two error sources add linearly.
The remaining errors are unrelated phenomena, so we assume they add in quadrature.
The final estimate of time variability is then \combinedNoise\ ppm per visit.
This indicates that the main factor impacting NIRCam grism time series will be correlated electronic noise, as discussed in \paperI\ and that the systematic noise floor will be less important.
The 1/f noise discussed in \paperI\ with the GRISMR disperser can be as high as 1000 ppm per integration so it remains the largest challenge to high precision.
Efforts to mitigate 1/f noise are underway, such as the GRISMC disperser being discussed as a science enhancement for the observatory.

We urge caution when assuming the low noise floor of \combinedNoise\ ppm, however.
Laboratory measurements are plagued by instabilities in light sources, where the current and temperature are difficult to control at the ppm level.
Of the laboratory tests measured at NASA centers and the University of Arizona laboratory, the most stable lightcurve had a standard deviation 320 ppm over 20 minutes after removing a linear trend.
The real noise floor will only be measured in flight on a more stable source such as a quiescent star while using the real flight configuration. 
There will be a commissioning exercise of an eclipsing binary to provide insight about the noise floor and stability of the NIRCam grism time series mode.

While we estimate a noise floor of \combinedNoise\ ppm and urge caution about unknown sources of error, corrections can be made to reduce the standard deviation of time series.
The error sources we discussed are potentially correctable by monitoring state variables of the instrument and telescope such as antenna moves, the image centroid, the charge trap state, detector temperature and the PSF shape.
Additionally, some of these noise sources (like small detector temperature fluctuations or IEC heaters) will be independent from one visit to the next.
We urge cautious estimates of the noise for the first JWST cycle but future cycles may achieve better performance.

We note that two noise sources can potentially produce a perennial offset in the transmission spectrum: the charge trapping effect and non-linearity.
The charge trapping, though small, will behave similarly from one visit to another and can introduce an exponential-like curvature to the time series.
Fortunately, this curvature is small (3 ppm) and will be correctable with algorithms such as RECTE \citep{zhou2017chargeTrap}.
Detector non-linearities and reciprocity failure will also occur in all visits because the illumination of the detectors will be the same.
Reciprocity failure can potentially introduce a $\sim$40 ppm spectral offset from the true value.
Detector non-linearities and reciprocity failure can potentially be calibrated with data from the absolute calibration program.
However, care must be taken when performing inter-instrument comparisons across JWST instruments or other observatories.

\added{
We note that the systematic errors discussed in this work are more broadly applicable to instruments and detectors beyond NIRCam and JWST.
The intrapixel sensitivity effects and crosshatching pattern discussed in Section \ref{sec:subpxCrosshatch} can occur on many different detectors.
Generally, the intrapixel sensitivity effects are most pronounced when a PSF is smaller than 2 pixels \citep[e.g. short wavelength Spitzer channels and JWST NIRSpec at short wavelengths][]{ingalls2016spitzerRepeatability,ferruit2014transitingPNIRSpec}, but less important when the PSF spans over many pixels \citep[e.g. JWST NIRISS][]{doyon2012NIRISSFGS}.
We therefore expect that NIRSpec prism/grating observations will have larger sub-pixel sensitivity than the estimated 6 ppm value estimated for NIRCam, but smaller for JWST NIRISS.
They will also be modified by the detector-dependent and spatially-dependent crosshatch pattern and any other sub-pixel detector structure.
Thus, the impact of crosshatching on observations that use HgCdTe detectors will depend on the optical design as well as the amplitude of the crosshatch pattern.

The variable aperture losses due to the telescope's thermal behavior discussed in Section \ref{sec:varApertureLosses} are also relevant for other space-based observatories.
For example, high precision time series observations are proposed for LUVOIR \citep{luvoir2019missionConceptReport}, Origins Space Telescope \citep{meixner2019ostReport} and the Nancy Grace Roman Telescope \citep{gaudi2019auxliaryScienceWFIRST}, which all could have long thermal settling timescales.
We expect the numbers in this paper to serve as useful starting points in estimating the noise in these observatories.
Finally, the crosshatching pattern, charge trapping effects, detector temperature fluctuations and non-linearity effects occur on all HgCdTe detectors used widely in astronomy \citep[e.g. WIRCam, HPF, ARCoIRIS, MOSFIRE, etc.][]{baril2006WIRCamH2RG,mahadevan2014hpf,schlawin2014TSpec,mclean2012mosfire}, so high precision measurements will be aided by image stability, temperature control and non-linearity corrections.
}

\acknowledgments

\section*{acknowledgements}
Thanks to Peiman Maghami for providing the line of site pointing jitter simulation and information on inter-boresight motion for use in estimating the subpixel sensitivity as well explaining sources of pointing error.
Thanks to Knicole Colon for useful discussions on the effect of pointing jitter on time series stability.
Thanks to Yifan Zhou for providing some charge trapping timescales for HST.
Rafia Bushra did extensive analysis of cryogenic vacuum testing that provided valuable insights about the various systematic errors that JWST can encounter.
Thanks to Sam Myers for bringing the \citet{suissa2020oceanEarthsMStars} paper to our attention.
Thanks to John Stansberry for providing information on the subarray field points and for tireless work to enable and improve the grism time series mode.
Thanks to Chaz Shapiro and Eric Huff for providing information on the Euclid detector's crosshatch pattern.
Funding for the E. Schlawin is provided by NASA Goddard Space Flight Center. 
T. Greene acknowledges funding from the NASA JWST project in WBS 411672.05.05.02.02.
\added{We thank the anonymous reviewer for their helpful suggestions to improve the paper and its readability.}

%

\vspace{5mm}
\facilities{JWST(NIRCam), NASA Goddard Space Environment Simulator, NASA Johnson Chamber A}


\software{\texttt{astropy} package \citep{astropy2013},
 \texttt{pynrc} \citep{leisenring2020pynrc0p8dev},
\texttt{webbpsf} \citep{perrin2014webbpsf},
\texttt{numpy} \citep{vanderWalt2011numpy},
\texttt{scipy} \citep{virtanen2020scipy},
\texttt{pysynphot} \citep{lim2015pysynphot},
\texttt{ncdhas},
\texttt{photutils} \citep{bradley2016photutilsv0p3},
\texttt{matplotlib} \citep{Hunter2007matplotlib}
}

\appendix

\section{Detailed Subpixel Crosshatching Characterization}\label{sec:detailedSupxCrosshatchingCharacterization}
\added{
Here, we primarily study the subpixel behavior of the A5 detector because it is used for for the NIRCam grism time series spectroscopy mode, covering 2.4\micron\ to 5.0\micron\ with a plate scale of 63 mas/px \citep{greene2017jatisNIRCam}.
The flat field of the A5 detector has a pronounced crosshatch pattern as shown in Figure~\ref{fig:crossHatchA5Zoom} and \ref{fig:crossHatchA5}.
The patterns are located at 23.1$\degree$ , 90.9$\degree$, and 158.6 $\degree$ counter-clockwise (CCW) from the $+$X direction of the detector.
The angles between these lines are 67.8$\degree$, 67.7$\degree$ and 44.5$\degree$.
These angles and patterns can be analyzed with a 2D power spectrum as shown in Figure \ref{fig:crossHatchA5}.

\begin{figure}[!hbtp]
\centering
\includegraphics[width=.49\columnwidth]{crosshatch_zoom.pdf}
\includegraphics[width=.49\columnwidth]{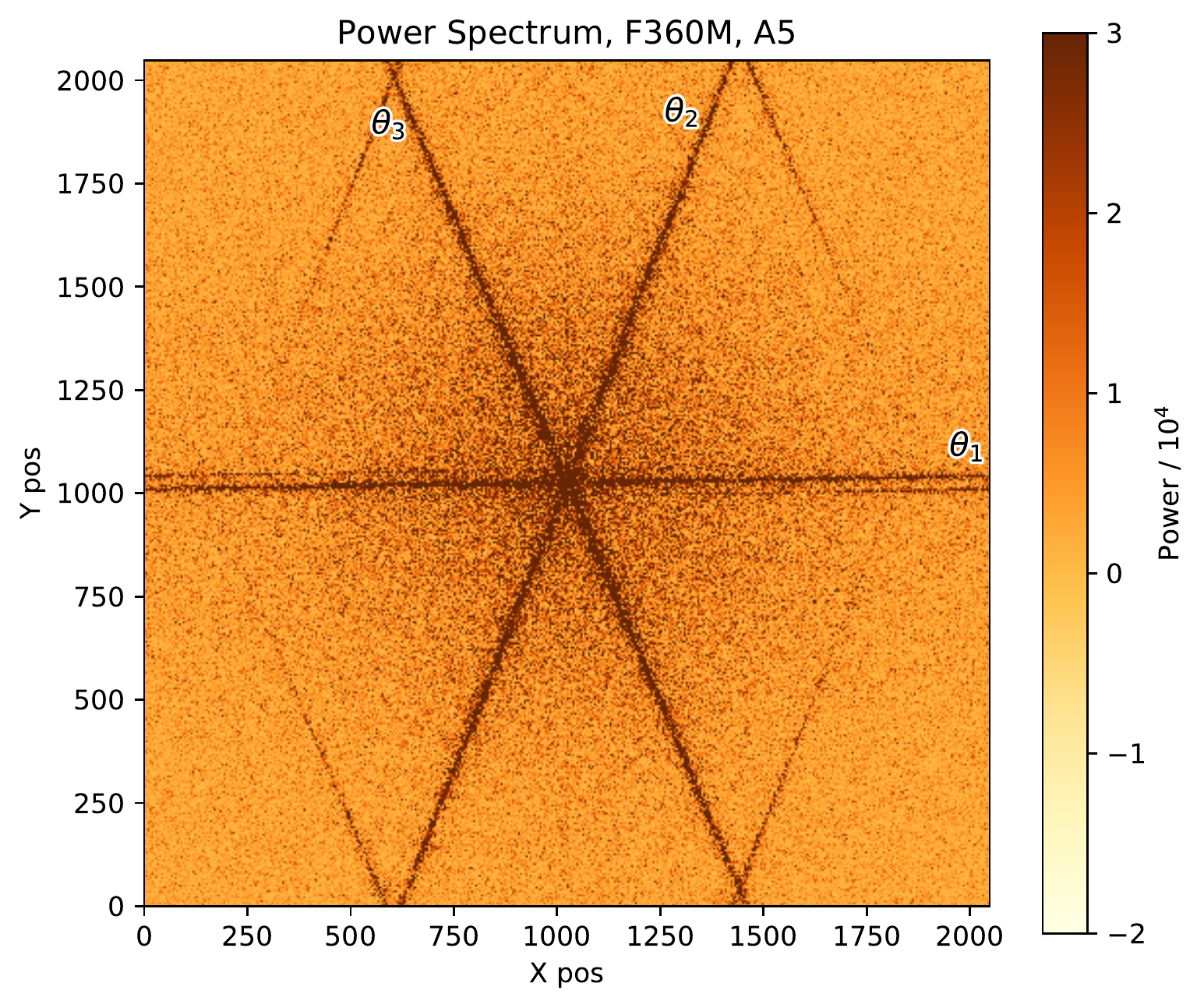}
\hspace{0.2in}
\includegraphics[width=.3\columnwidth]{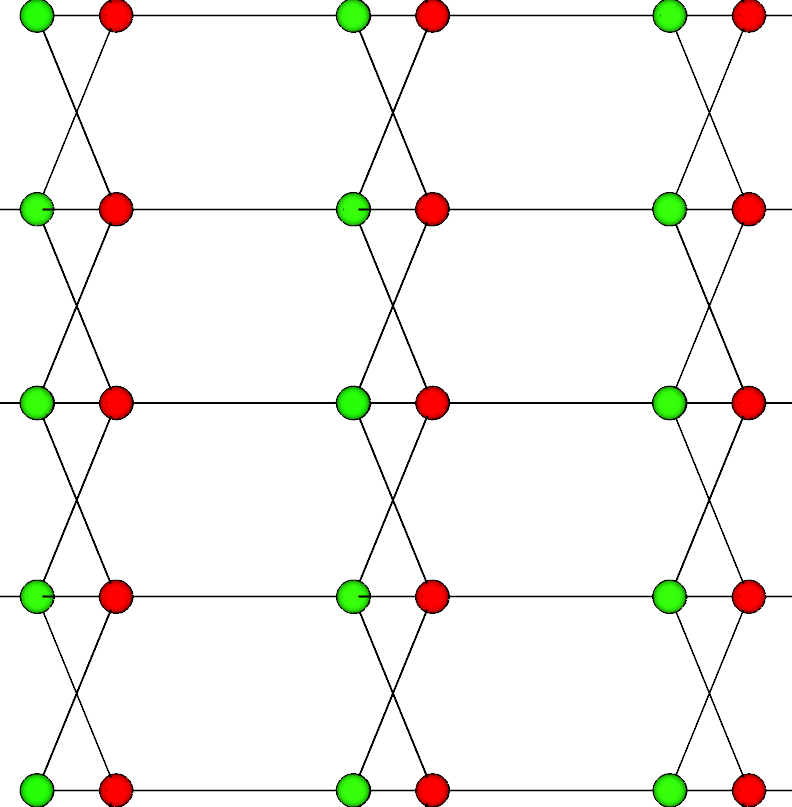}
\caption{
\added{The A5 detector's flat field has a pronounced crosshatch pattern at $\alpha_1 = 90.9 \degree$, $\alpha_2 = 158.6\degree$ and $\alpha_3 =23.1\degree$ CCW from the positive X direction (top left plot).
The flat field from Figure \ref{fig:crossHatchA5Zoom} is reproduced here.
The crosshatch pattern is best analyzed with a 2D Fourier power spectrum (top right), which shows a continuum of frequencies aligned with the three crosshatch angles.
Note that the lines in the frequency plane ($\theta_i$) are perpendicular to the crosshatch directions in the image plane ($\alpha_i$).
These three relative crosshatch angles (67.8$\degree$, 67.7$\degree$ and 44.5$\degree$) are similar to a projection of a tetrahedral (zincblende) lattice structure of HgCdTe (bottom), 67.8$\degree$, 67.8$\degree$, 44.4$\degree$).
The green circles represent either Cd or Hg while the red represents Te.}
}\label{fig:crossHatchA5}
\end{figure}

The angles of the crosshatch patterns are determined by the crystal pattern of HgCdTe.
HgCdTe has a zincblende structure with tetrahedral bond angles where each Hg or Cd atom is surrounded by 4 Te atoms \citep{gemain2012mercVacanciesHgCdTe}.
HgCdTe detectors are manufactured using molecular beam epitaxy upon a substrate, a process which can result in topological defects with peak-to-valley amplitudes of 5-20~nm in height variations \citep{chang2008surfaceMorphologyHgCdTe}.

The surface morphology of the HgCdTe crystal shows that the crosshatch patterns are oriented along the intersection of the (211) growth plane of the crystal and the 8 HgCdTe slip planes.
The relative angles of the HgCdTe slip planes and (211) growth plane are 44.42$\degree$, 67.79$\degree$ and 67.79$\degree$ \citep{chang2008surfaceMorphologyHgCdTe}, very close to the observed crosshatch angles.
A projection of zincblende structure is shown in Figure~\ref{fig:crossHatchA5} at the same orientation of the power spectrum structure.
The similarity between the crosshatch patterns in the flat field and the topological variations observed in \citet{chang2008surfaceMorphologyHgCdTe} leads to the likely conclusion that the surface variations lead to quantum efficiency variations.
Thus, the crosshatch pattern is most likely related to the crystal lattice structure of the HgCdTe substrate and not the pixel circuitry or readout electronics.

The crosshatch structure of the detectors extends down to the subpixel level, so it will not be fully corrected with a flat field division.
This sub-pixel structure has been imaged with microscopy on a candidate Euclid HgCdTe detector \citep{shapiro2018crosshatch}.
Atomic force microscopy shows topological features that are approximately 1.2~$\mu$m in width \citep{chang2008surfaceMorphologyHgCdTe} compared to the 18~$\mu$m pixel sizes.

The width of the structures can also be estimated from the behavior for crosshatch lines as they cross pixel boundaries \citep{ninan2019crosshatchHPF}.
We estimate that the crosshatch pattern oriented at 90.9$\degree$ crosses 1 horizontal pixel for every 64 vertical pixels.
While crossing the boundary, there are $\approx 38$ rows where a sharp feature in the crosshatch pattern spans 2 pixels instead of 1.
The sharp feature is traced by the pixels that drop significantly below the mean flat field value in a local region.
If the sharp feature in the crosshatch pattern has a tophat shape, then these 38 rows where the pattern spans two pixels imply a tophat full width of 0.6 pixels or a physical width of 10.8~$\mu$m for an 18~$\mu$m pixel pitch.
This is more than twice the estimate from \citet{ninan2019crosshatchHPF} for an HgCdTe used on the Habitable Zone Planet Finder (HPF) instrument.
We expect that the crosshatch pattern's width varies among detectors or that a tophat function is a poor approximation of the actual subpixel response of this detector.
Within NIRCam detectors, there are large variations in the strength and orientation of crosshatch features.

The crosshatch pattern is wavelength dependent, as seen in Figure~\ref{fig:crossHatchWavelengthDep}.
Here, the flat field for the A5 detector as well as the 2D Fourier Amplitude are shown, which is the same data used in Figure \ref{fig:crossHatchA5}.
The throughput variations are the largest for short wavelengths (better resolving crystal structures) and smallest for the long wavelengths.
This is visible both in the throughput cross section of the flood-illuminated flat field and the Fourier amplitude of the crosshatch as a function of wavelength.
We find a steeper wavelength dependence to the crosshatch pattern near 0.9$\degree$ from horizontal in the frequency domain or 0.9$\degree$ from vertical in the length domain.

\begin{figure}[!hbtp]
\centering
\includegraphics[width=.49\columnwidth]{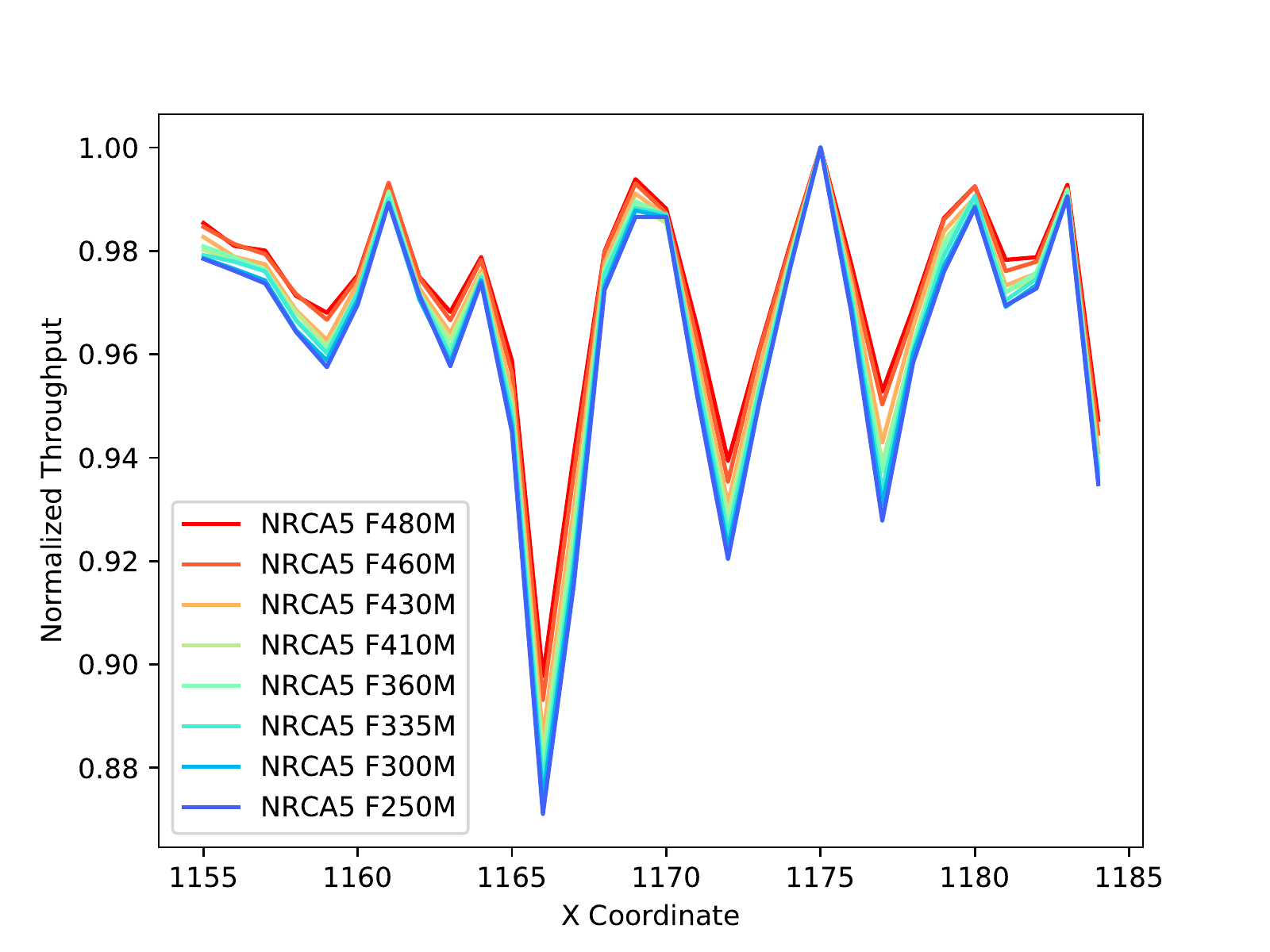}
\includegraphics[width=.49\columnwidth]{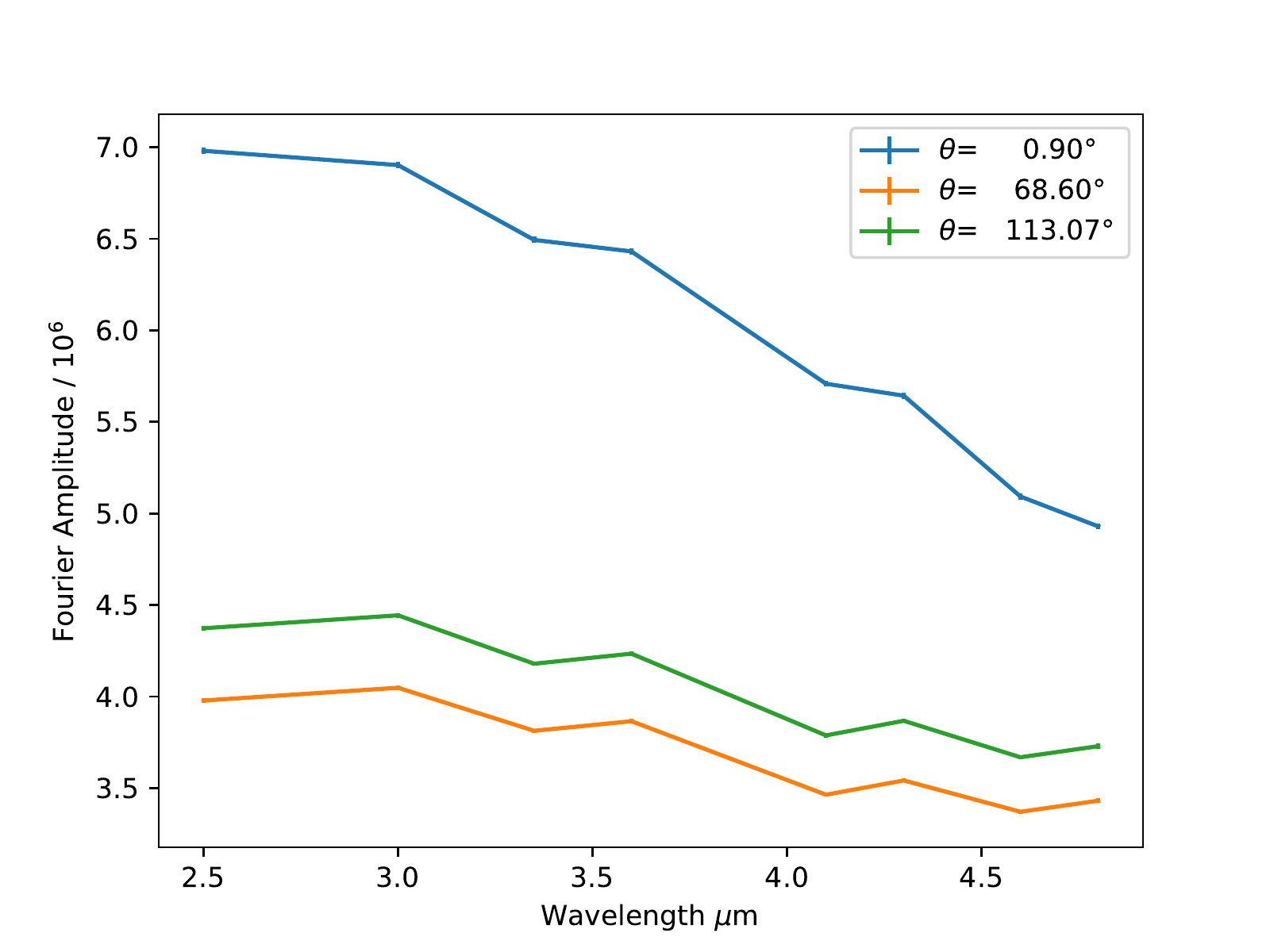}
\caption{
The crosshatch pattern changes as a function of wavelength, likely because the shorter wavelengths resolve the structure better.
In other words, the crosshatch pattern becomes slightly ``blurred'' at longer wavelengths due to diffraction.
The shorter wavelengths have deeper crosshatch troughs in a flood-illuminated flat field (left).
The flat fields are shown here for the A5 detector, the same one shown in Figure \ref{fig:crossHatchA5}.
The sharper crosshatch features at shorter wavelengths show up as larger Fourier amplitudes (right).
}\label{fig:crossHatchWavelengthDep}
\end{figure}
}

\section{Map of Regions}\label{sec:regionMap}
\begin{figure}[!hbtp]
\centering
\includegraphics[width=.79\columnwidth]{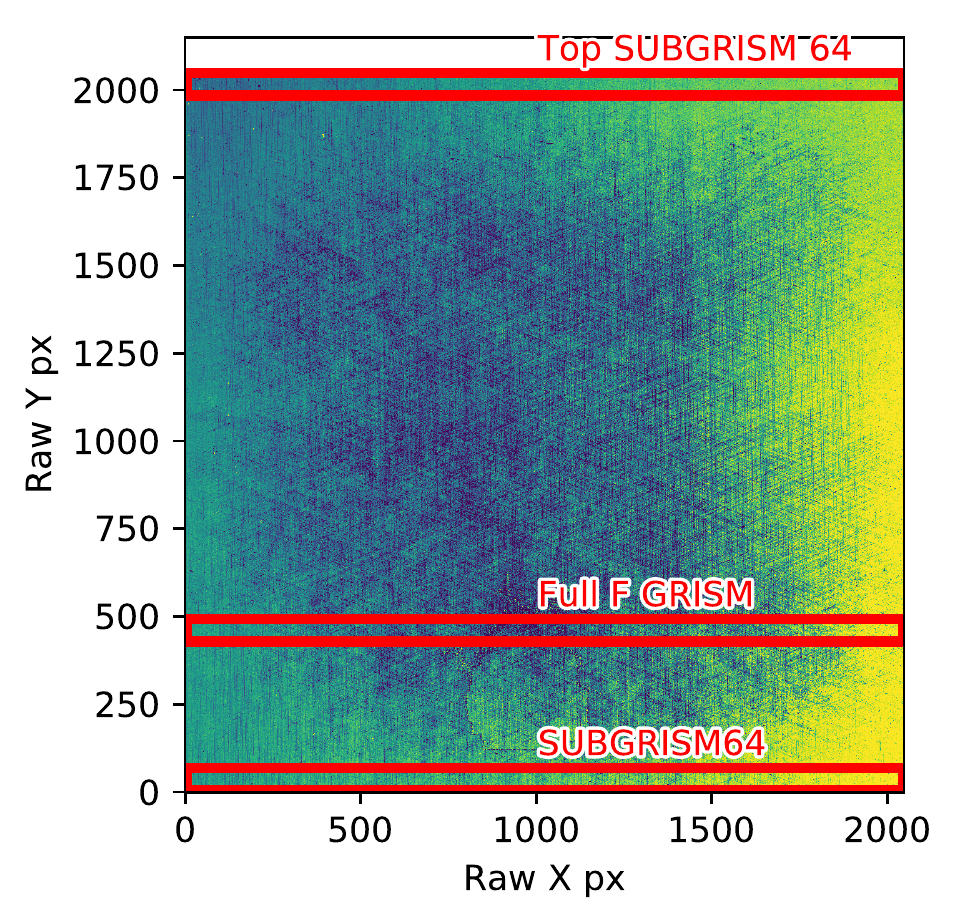}
\caption{The three regions considered in Section \ref{sec:subgrismPos} are shown on a full frame flat field image of the A5 detector.}\label{fig:regionMap}
\end{figure}

In Section \ref{sec:subgrismPos}, we describe three regions on the A5 detector where one could place a dispersed grism image with varying degrees of crosshatch amplitude.
Figure \ref{fig:regionMap} provides a map of those three regions on top of the flat field image for reference.

\section{Charge Trapping Schematic}\label{sec:chargeTrapSchematic}
All HgCdTe detectors show a signal after bright illumination even if the illumination is removed; this is called persistence.
The physical mechanism that explains persistence is that charges are trapped in the depletion region after illumination by photons as the detector well fills \citep{smith2008imgPersistence}.
This was specifically measured for the NIRCam detectors in \citet{leisenring2016persistence}.
After the reverse-bias voltage is applied (a reset) in an ideal detector, the depletion region will be devoid of mobile charge carriers (electrons and holes).
However, in real detectors, some charges are trapped within the depletion region.
During a future integration, these charges will be released and shrink the size of the depletion region, recording a spurious signal (Data Numbers) for that pixel not related to the current integration's external illumination.

\begin{figure}[!hbtp]
\centering
\includegraphics[width=.99\columnwidth]{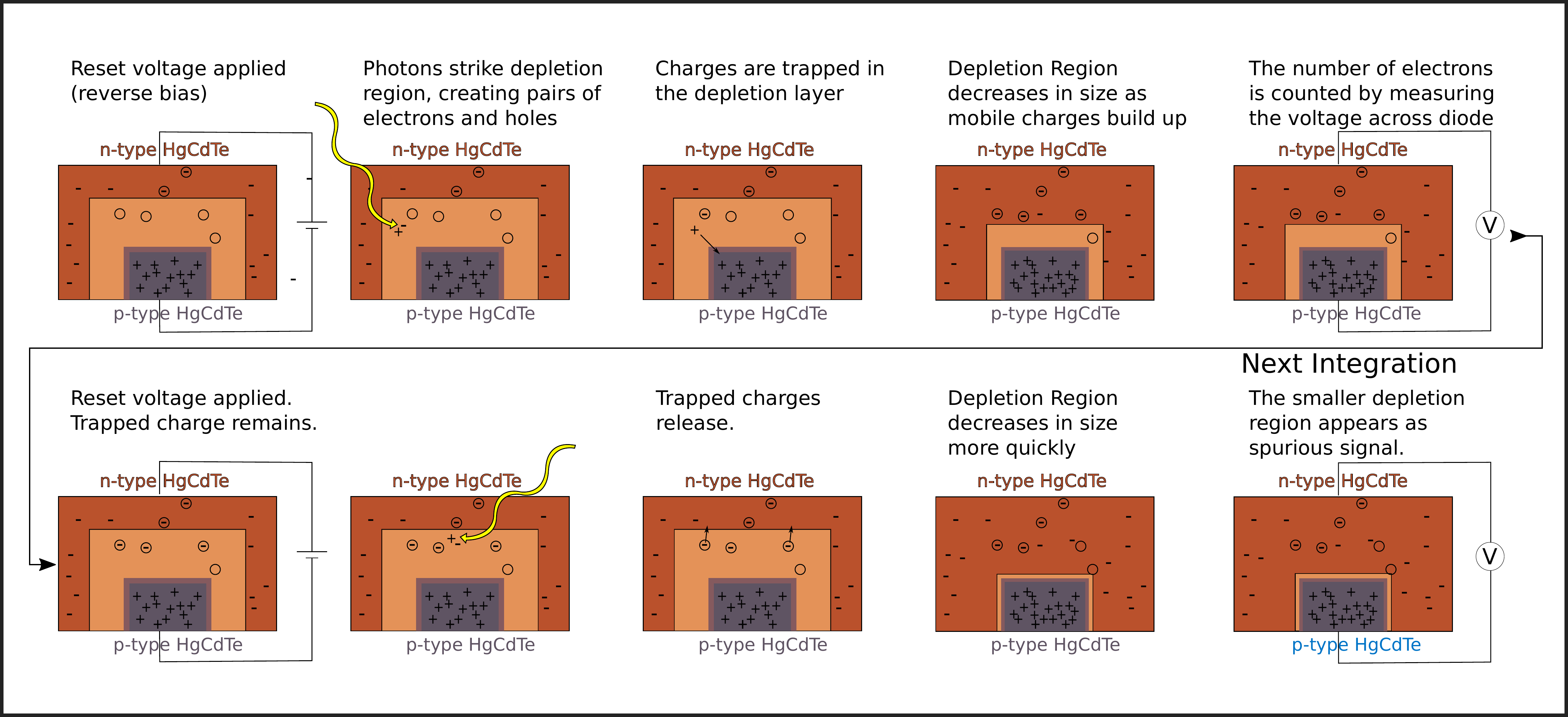}
\caption{A cartoon schematic of a H2RG detector with charge traps inspired by \citet{smith2008imgPersistence}, \citet{tulloch2018persistenceH2RG}, and \citet{leisenring2016persistence}.
Charged traps for negative and positive carriers (depicted as round circles) will capture charge before it can flow to the undepleted regions of the detector.
The charge traps do not immediately empty with a detector reset, but instead release at later times causing a spurious signal.
This spurious signal (the shrinking of the depletion region), causes a voltage change in future integrations that appears the same as a true signal from incoming photons.}\label{fig:npSchematicTraps}
\end{figure}

Figure~\ref{fig:npSchematicTraps} shows a schematic of a detector that has charge traps in the substrate.
After illumination by a source, these traps will fill with charge that does not migrate across the depletion layer.
When the reset voltage is applied to the detector, the trapped charge remains within the depletion layer.
In the second integration in the schematic, charge is released, which reduces the size of the depletion layer faster than the incoming photons would otherwise.
The consequence of charge traps is that there is a deficiency in measured charge on the first integration and an excess of measured charge on the second integration.
Therefore, a constant astrophysical signal can be measured as a time-varying one due to the charge trapping effect.
Over timescales longer than the charge release time, the detector will reach a steady state in the latent signal's functional form assuming the observed scene stays constant.
As discussed in Section \ref{sec:persistence}, the flight NIRCam detector shows low levels of persistence compared to previous generation detectors used in HST.

\bibliographystyle{apj}
\bibliography{this_biblio}



\end{document}